\documentclass[12pt,english,onecolumn, draftcls, conference]{IEEEtran}

\usepackage[T1]{fontenc}
\usepackage[latin1]{inputenc}
\usepackage{geometry}
\usepackage{amsthm}
\usepackage{amsmath}
\usepackage{amssymb}
\usepackage{graphicx}
\usepackage{epstopdf}
\usepackage{cite}
\usepackage{color}
\usepackage{mathtools}
\usepackage{cases}
\usepackage{stackengine}
\usepackage{accents}

\newcommand\blfootnote[1]{%
  \begingroup
  \renewcommand\thefootnote{}\footnote{#1}%
  \addtocounter{footnote}{-1}%
  \endgroup
}

\newtheorem{thm}{Theorem}[section]

\newtheorem{prop}[thm]{Proposition}

\newtheorem{rem}{Remark}[section]

\renewcommand\appendix{\par
\setcounter{section}{0}
\setcounter{subsection}{0}
\setcounter{figure}{0}
\setcounter{table}{0}
\renewcommand\thesection{ \Alph{section}}
\renewcommand\thefigure{\Alph{section}\arabic{figure}}
\renewcommand\thetable{\Alph{section}\arabic{table}}
}

\usepackage{babel}

\allowdisplaybreaks

\begin{document}
\title{Zero-Delay Source-Channel Coding with a One-Bit ADC Front End and Correlated Side Information at the Receiver}
\author{Morteza Varasteh$^\dagger$, Borzoo Rassouli$^\dagger$, Osvaldo Simeone$^*$ and Deniz G\"{u}nd\"{u}z$^\dagger$ \\
$^\dagger$ Department of Electrical and Electronic Engineering, Imperial College London, London, U.K.\\
$^*$ CWCSPR, ECE Department, New Jersey Institute of Technology, NJ, USA.\\
\{m.varasteh12; b.rassouli12; d.gunduz\}@imperial.ac.uk, osvaldo.simeone@njit.edu. }

\maketitle
\begin{abstract}
Zero-delay transmission of a Gaussian source over an additive white Gaussian noise (AWGN) channel is considered with a one-bit analog-to-digital converter (ADC) front end and a correlated side information at the receiver. The design of the optimal encoder and decoder is studied for two performance criteria, namely, the mean squared error (MSE) distortion and the distortion outage probability (DOP), under an average power constraint on the channel input. For both criteria, necessary optimality conditions for the encoder and the decoder are derived. Using these conditions, it is observed that the numerically optimized encoder (NOE) under the MSE distortion criterion is periodic, and its period increases with the correlation between the source and the receiver side information. For the DOP, it is instead seen that the NOE mappings periodically acquire positive and negative values, which decay to zero with increasing source magnitude, and the interval over which the mapping takes non-zero values, becomes wider with the correlation between the source and the side information.\blfootnote{This work was presented in part at the 2016 IEEE Information theory workshop \cite{ITW_link_2016}.

M. Varasteh and B. Rassouli have been supported by the British Council Institutional Links Program under grant number 173605884.

D. Gunduz has received funding from the European Research Council (ERC) through Starting Grant BEACON (agreement No. 677854).

The work of O. Simeone has received funding from the European Research Council (ERC) under the European Union's Horizon 2020 research and innovation programme (grant agreement No 725731), and was partially supported by the U.S. NSF through grant CCF-1525629.}

\textit{\textbf{Index Terms}}- Joint source channel coding, zero-delay transmission, mean squared error distortion, distortion outage probability, one-bit ADC, correlated side information.
\end{abstract}

\section{Introduction}\label{Chptr4_Sec:Introduction}
Current wireless communication systems enable reliable transmission of specific high-rate content types, such as JPEG and MPEG, by exploiting near capacity-achieving channel codes and highly optimized compression algorithms. However, many emerging applications, such as the Internet-of-Things (IoT) or machine-to-machine (M2M) communications, impose further constraints on the cost and complexity of communication devices, or on the available energy and the end-to-end latency, which render many of the known codes and modulation techniques inapplicable. For example, in time-sensitive control applications, such as the monitoring of power lines for attacks or failures in a smart grid, or detection and prevention of epileptic seizures through embedded sensors, the underlying signals should be measured and transmitted to the receiving-end under extreme latency constraints. In such scenarios, neither measuring multiple signals to improve the compression efficiency, nor using the channel many times to approach the channel capacity is possible. Here, we model such a communication scenario with the extreme \textit{zero-delay} constraint, imposing the transmission of a single sample of the underlying signal over a single use of the channel.

A key component of the front end of any digital receiver is the analog-to-digital converter (ADC) that is typically connected to each receiving antenna. The energy consumption of an ADC (in Joules/sample) increases exponentially with its resolution (in bits/sample) \cite{Murmann_2014}. This is causing a growing concern regarding the energy consumption of digital receivers, either due to the increasing number of receiving antennas, e.g., for massive multiple-input multiple-output (MIMO) transceivers \cite{Risi_Persson_Larsson_2014}, or due to the limited availability of energy, e.g., in energy harvesting terminals \cite{Gunduz_Stamatiou_Michelusi_Zorzi_2014}. Energy-efficient operation of digital receivers may hence impose constraints on the resolution of the ADCs that can be employed for each receiving antenna.

\begin{figure}
\begin{centering}
\includegraphics[scale=0.95]{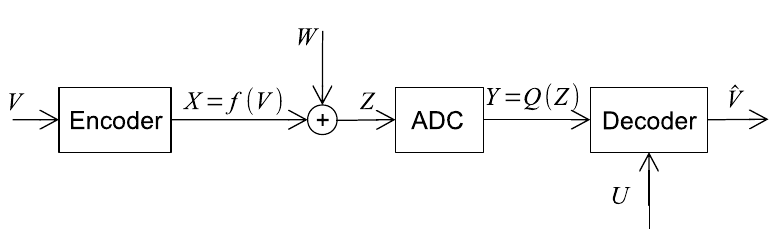}
\caption{System model for the zero-delay transmission of a Gaussian source sample over an AWGN channel with a one-bit ADC front end and correlated side information at the receiver.}\label{Chptr4_Figure:1}
\par\end{centering}
\vspace{0mm}
\end{figure}

Motivated by communication among energy- and complexity-limited sensor nodes under extreme latency constraints, we study the zero-delay transmission of analog sensor measurements to a receiver equipped with a 1-bit ADC front end. Considering that the transmitter and the receiver should be physically close to each other when communicating at low power, we further assume that the receiving node has its own correlated measurement of the transmitted source sample (see Figure \ref{Chptr4_Figure:1}). We consider two standard performance criteria, namely the mean squared error (MSE) distortion and the distortion outage probability (DOP). Our goal here is to gain insights into the structure and the performance of the optimal encoder and decoder functions when the source sample and the side information are jointly Gaussian.

This work contributes to a line of research that endeavors to understand the impact of front-end ADC limitations on the fundamental performance limits of communication systems. The capacity analysis of a real discrete-time AWGN channel with a $K$-level ADC front end is studied in \cite{Singh_Dabeer_Madhow_2009}, proving the sufficiency of $K+1$ constellation points at the encoder. Furthermore, it is shown in \cite{Singh_Dabeer_Madhow_2009} that BPSK modulation achieves the capacity when the receiver front end is limited to a 1-bit ADC. In \cite{Koch_Lapidoth_2013}, the authors prove that, in the low signal-to-noise ratio (SNR) regime, the symmetric threshold 1-bit ADC is suboptimal, while asymmetric threshold quantizers and asymmetric signalling constellations are needed to obtain the optimal performance. The generalization of the analysis to multiple-input multiple-output (MIMO) fading systems is put forth in \cite{Mezghani_2007}, and, more recently, to massive MIMO systems in \cite{Risi_Persson_Larsson_2014} and \cite{Jacobsson_Durisi_Coldrey_Gustavsson_Studer_2015}. In \cite{arive_varsteh_rassouli_osvaldo_gunduz} the authors of this work considered the zero-delay transmission set-up analysed here, but in the absence of correlated side information at the receiver. It is noted that the zero-delay constraint prevents the application of the channel capacity results in \cite{Singh_Dabeer_Madhow_2009,Jacobsson_Durisi_Coldrey_Gustavsson_Studer_2015}, and as it will be seen, the presence of correlated side information at the receiver significantly modifies the optimal design problem with respect to the set-up studied in \cite{arive_varsteh_rassouli_osvaldo_gunduz}.

The main contributions of this work are as follows. We derive necessary optimality conditions for encoder and decoder mappings for both of the performance criteria under consideration, namely the MSE and the DOP. Based on these conditions, for the MSE criterion, we observe that, in a manner similar to the case with an infinite resolution front end studied in \cite{Akyol_Viswanatha_Rose_2014, Mehmetoglu_Akyol_Rose_2013, Chen_Tuncel_2011}, the numerically optimized encoder (NOE) mapping is periodic. Furthermore, the period of this function depends on the correlation coefficient between the source and the side information, and is independent of the input power constraint, or equivalently the channel SNR. Motivated by the structure of the NOE mappings, we also propose two simple parameterized mappings, which, although being suboptimal, approach the performance of NOE mappings in the low- and high-SNR regimes. For the DOP criterion, we also observe that the NOE mappings periodically acquire positive and negative values. Additionally, the NOE mappings for the DOP criterion decay to zero with increasing source magnitude. It is also observed that, as the correlation between the source and the side information increases, the number of changes between positive and negative values in the encoder mapping, as well as the size of the intervals of source output values for which the encoder mapping is non-zero, increase.

The rest of the paper is organized as follows. In Section \ref{Chptr4_Sec:System Model}, the system model is explained. Section \ref{Chptr4:Sec_Average Distortion} focuses on the MSE criterion. We study the optimal design of the encoder and the decoder in Section \ref{Chptr4:Sec_Optimal_design}, while Sections \ref{Chptr4:Sec_linear} and \ref{Chptr4:Sec_digital} present two baseline suboptimal encoding schemes. In Section \ref{Chptr4:Sec_preresult}, we consider the scenario in which the side information is also available at the encoder, and, by leveraging the results in \cite{arive_varsteh_rassouli_osvaldo_gunduz}, we obtain a lower bound on the performance of the original problem with decoder-only side information. As another reference result, in Section \ref{Chptr4:Sec_Lowerbound}, we present the Shannon lower bound for the decoder-only side information problem. Focusing on the DOP criterion in Section \ref{Chptr4:Sec_DOP}, we first consider the optimal design of the encoder and decoder in Section \ref{Chptr4:Sec_Outage_Distortion}. Next, in Section \ref{Chptr4:Sec_Previous Result}, as for the MSE counterpart, we consider the case in which the side information is also available at the encoder. In Section \ref{Chptr4:Sec_Numerical Results}, numerical results are provided, and Section \ref{Chptr4:Sec_conclusion} concludes the paper.

\textit{Notations}: Throughout the paper upper case and lowercase letters denote random variables and their realizations, respectively. The standard normal distribution is denoted by $\mathcal{N}(0,1)$, and its probability density function (pdf) by $\Phi(\cdot)$. $\mathbb{E}[\cdot]$ and $\textrm{Pr}(\cdot)$ stand for the expectation and probability, respectively. $Q(\cdot)$ denotes the complementary cumulative distribution function (CCDF) of the standard normal distribution, defined as
\begin{align}
Q(z)\triangleq\frac{1}{\sqrt{2\pi}}\int\limits_{z}^{\infty}e^{-\frac{x^2}{2}}dx.
\end{align}
The boundaries of integrals are from $-\infty$ to $\infty$ unless stated otherwise. We denote the pdf of a standard bivariate normal distribution with correlation $r$ as
\begin{align}
\Phi\left(v,u\right)&=\frac{1}{2\pi\sqrt{1-r^2}}
e^{-\frac{1}{2(1-r^2)}\left(v^2+u^2-2 r vu\right)},
\end{align}
and the conditional pdf for these variables as
\begin{align}
\Phi\left(v|u\right)&=\frac{1}{\sqrt{2\pi(1-r^2)}}
e^{-\frac{\left(v-ru\right)^2}{2(1-r^2)}}.
\end{align}

\vspace{5mm}

\section{System Model}\label{Chptr4_Sec:System Model}
We consider the system model in Figure \ref{Chptr4_Figure:1}, in which a single Gaussian source sample $V\sim \mathcal{N}(0,\sigma_v^2)$ is transmitted over a single use of a channel characterized by AWGN followed by a one-bit ADC front end. Unlike the model studied in \cite{arive_varsteh_rassouli_osvaldo_gunduz}, the receiver has access to side information $U\sim \mathcal{N}(0,\sigma_u^2)$, which is correlated with the source $V$. The correlation matrix of the source and the side information is given by
\begin{equation}
\Lambda=\left[
          \begin{array}{cc}
            \sigma^2_v & r \sigma_v\sigma_u \\
            r\sigma_v\sigma_u & \sigma_u^2 \\
          \end{array}
        \right],
\end{equation}
where $r\in[-1,1]$ denotes the correlation coefficient.

The encoded signal is obtained as $X=f(V)$, where $f:\mathbb{R} \rightarrow \mathbb{R} $ is a mapping that is constrained to satisfy an average power constraint $\mathbb{E}[f(V)^2]\leq P$. At the receiver, the received noisy signal is modelled as
\begin{equation}
Z=f(V)+W,
\end{equation}
where $W\sim \mathcal{N}(0,\sigma_w^2)$ is independent of the source and side information. The noisy signal $Z$ is quantized with a one-bit ADC producing the received signal as
\begin{equation}\label{Chptr4:Eqn_1}
Y=\mathcal{Q}(Z)=\left\{\begin{array}{ll}0 & Z \geq 0, \\ 1 & Z < 0. \end{array}\right.
\end{equation}
We define the signal-to-noise ratio (SNR) as $\gamma=P/\sigma_w^2$. Based on $Y$ and $U$, the decoder produces an estimate $\hat{V}$ of $V$ using a decoding function $g:\{0,1\}\times \mathbb{R}\rightarrow \mathbb{R}$, i.e., $\hat{V}=g(Y,U)$.

Two performance criteria are considered in this paper, namely, the MSE distortion
\begin{equation}
\bar{D}=\mathbb{E}\left[(V-\hat{V})^2\right],
\end{equation}
and the DOP
\begin{equation}\label{Chptr4:Eqn_45}
\epsilon(D)=\mathrm{Pr}\left((V-\hat{V})^2\geq D\right).
\end{equation}
In both cases, we aim at studying the optimal encoder mapping $f$, along with the corresponding optimal estimator $g$ at the receiver, such that $\bar{D}$ and $\epsilon(D)$ are minimized subject to the average power constraint. More specifically, as it is common in related works (see, e.g., \cite{Akyol_Viswanatha_Rose_2014}), we consider the unconstrained minimization
\begin{equation}\label{Chptr4:Eqn_2}
\begin{aligned}
& \underset{f,g}{\text{minimize}}
& & L(f,g,\lambda),
\end{aligned}
\end{equation}
where
\begin{align}\label{Chptr4:Eqn_3}
  L(f,g,\lambda) &=\left\{\begin{array}{ll}
                   \!\!\!\bar{D}+\lambda E[f(V)^2] & \text{for the MSE criterion} ,\\
                   \!\!\!\epsilon(D)+\lambda E[f(V)^2] & \text{for the DOP criterion},
                 \end{array}\right.\!\!\!
\end{align}
with $\lambda\geq 0$ being a Lagrange multiplier that defines the relative weight given to the average transmission power $\mathbb{E}[f(V)^2]$ as compared to the distortion criterion.

\section{MSE distortion criterion}\label{Chptr4:Sec_Average Distortion}
In this section, we study the performance of the system model in Figure \ref{Chptr4_Figure:1} under the MSE distortion criterion. In the following, we first consider the optimal design of the encoder and the decoder, and obtain a necessary condition for the optimality of an encoder mapping. For reference, we also study two parameterized encoding schemes, namely periodic linear transmission (PLT) and periodic BPSK transmission (PBT). Then, as lower bounds, we consider the MSE in the presence of side information at both the encoder and the decoder, and the Shannon lower bound.

\subsection{Optimal Encoder and Decoder Design}\label{Chptr4:Sec_Optimal_design}
 The design goal is to minimize the Lagrangian in (\ref{Chptr4:Eqn_2}) for the MSE distortion criterion. For any encoding function, the optimal decoder is the minimum MSE (MMSE) estimator
\begin{subequations}\label{Chptr4:Eqn_4}
\begin{align}
\hat{v}=g(y,u)&=\mathbb{E}[V|Y=y,U=u]\\
&=\frac{\int v\Phi\left(\frac{v}{\sigma_v}\big|\frac{u}{\sigma_u}\right)Q\left(\frac{(-1)^{y+1}f(v)}{\sigma_w}\right)dv}
{\int\Phi\left(\frac{v}{\sigma_v}\big|\frac{u}{\sigma_u}\right)Q\left(\frac{(-1)^{y+1}f(v)}{\sigma_w}\right)dv}.
\end{align}
\end{subequations}
The following proposition provides a necessary condition for the optimal encoder mapping.

\begin{prop}\label{Chptr4:Prop_1}
The optimal encoder mapping $f$ for problem (\ref{Chptr4:Eqn_2}) under the MSE distortion criterion must satisfy the implicit equation
\begin{align}\label{Chptr4:Eqn_5}
2\sqrt{2\pi}\sigma_w\sigma_u\lambda f(v)e^{\frac{f(v)^2}{2\sigma_w^2}}&=2v A(v)-B(v),
\end{align}
where $\lambda\geq 0$ and is given. The functions $A(v)$ and $B(v)$ are defined as
\begin{subequations}
\begin{align}
A(v)&\triangleq\int \Phi\left(\frac{u}{\sigma_u}\Big|\frac{v}{\sigma_v}\right)\left(g(0,u)-g(1,u)\right)du,\\
B(v)&\triangleq\int \Phi\left(\frac{u}{\sigma_u}\Big|\frac{v}{\sigma_v}\right)\left(g(0,u)^2-g(1,u)^2\right)du,
\end{align}
\end{subequations}
and $g(y,u)$, for $y=0,1$, is the optimal MMSE estimator defined in (\ref{Chptr4:Eqn_4}). Furthermore, the gradient of the Lagrangian function $L(f,g,\lambda)$ over $f$, for $g$ given as in (\ref{Chptr4:Eqn_4}), is given by
\begin{align}\label{Chptr4:Eqn_39}
\nabla L&=2\lambda f(v)-\frac{e^{-\frac{f(v)^2}{2\sigma_w^2}}}{\sqrt{2\pi}\sigma_w\sigma_u}(2vA(v)-B(v)).
\end{align}
\end{prop}
\textit{Proof}: See Appendix \ref{Chptr4:Appendix_2}.

\begin{figure}
\begin{centering}
\includegraphics[scale=0.5]{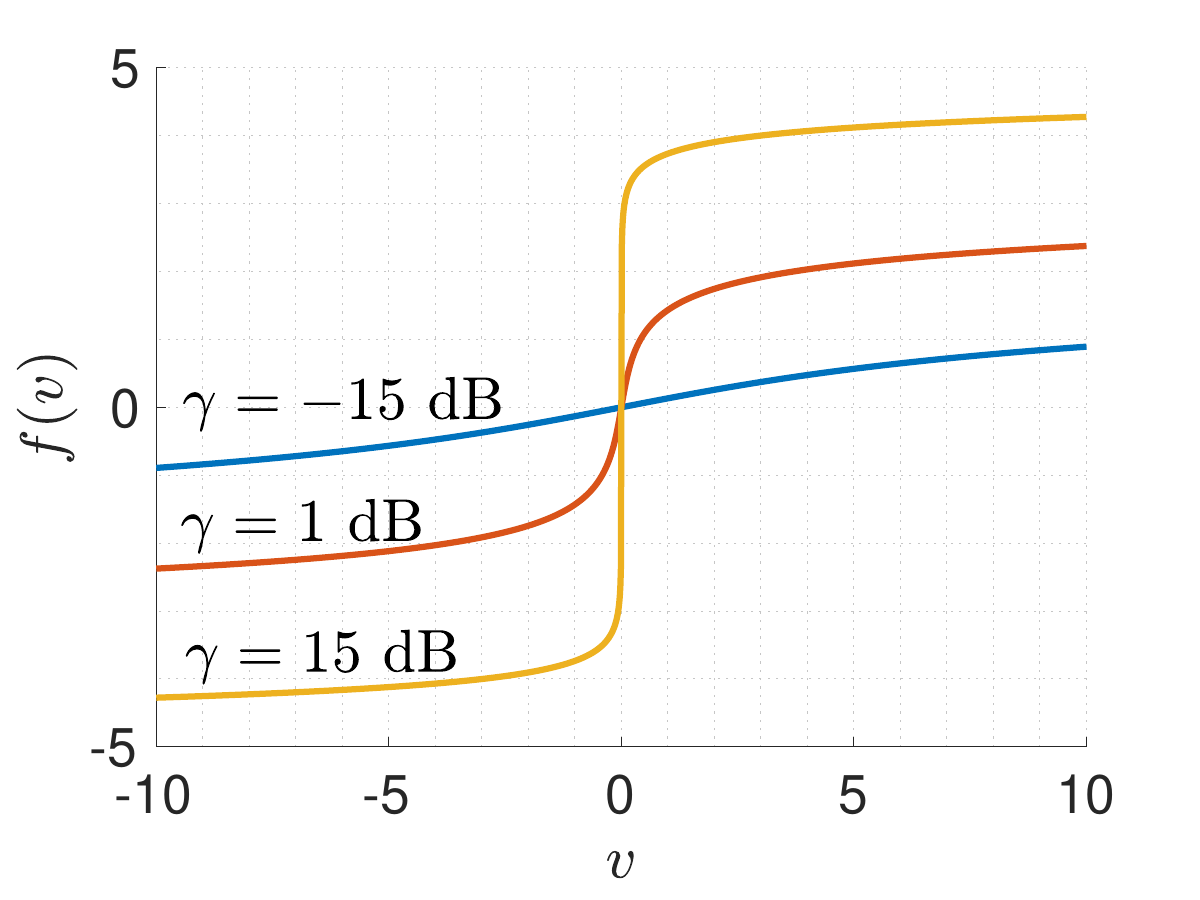}
\caption{Illustration of the optimal encoder mapping when there is no side information at the receiver, i.e., $r=0$ ($\sigma_v^2=\sigma_w^2=1$).}\label{Chptr4:Figure_4}
\par\end{centering}
\vspace{0mm}
\end{figure}

To elaborate on the necessary condition obtained in (\ref{Chptr4:Eqn_5}), we consider two extreme values of the correlation coefficient $r$. If we have an independent side information, that is, when $r=0$, it can be easily verified that the condition (\ref{Chptr4:Eqn_5}) coincides with the result obtained in \cite[Proposition III.1]{arive_varsteh_rassouli_osvaldo_gunduz} without considering a side information at the receiver. The optimal mapping in this case is an odd function. Plot of the optimal encoder mapping for different SNR values is shown in Figure \ref{Chptr4:Figure_4}. It can be seen that, for high SNR (large $\gamma$), the mapping tends to binary antipodal signalling, whereas for low SNR (small $\gamma$), it tends to a linear mapping. In contrast, with perfect side information, i.e., $r=\pm1$, we have $\Phi\left(u/\sigma_u|v/\sigma_v\right)=\pm\frac{\sigma_u}{\sigma_v}\delta(u-v)$,
where $\delta(\cdot)$ is the Dirac delta function, and, from (\ref{Chptr4:Eqn_5}), it is seen that the optimal mapping
is $f(v)=0$. Therefore, $g(y,u)$ in (\ref{Chptr4:Eqn_4}) is the MMSE estimate of $V$ given $U$, namely $g(y,u)=\frac{\sigma_v}{\sigma_u}u$.

In Section \ref{Chptr4:Sec_Numerical Results}, we will present NOE mappings obtained via gradient descent by using (\ref{Chptr4:Eqn_39}). We will observe that, due to the correlated receiver side information, the resulting encoder mappings are periodic, with a period that depends on the correlation coefficient $r$. Similar periodic mappings have been found to be optimal in \cite{Akyol_Viswanatha_Rose_2014} for the case with an infinite-resolution front end. Motivated by this observation and by the results in \cite{arive_varsteh_rassouli_osvaldo_gunduz} for the case of no side information (see Figure \ref{Chptr4:Figure_4}), we now propose two simple parameterized encoder mappings, which will be compared with the NOE mapping in Section \ref{Chptr4:Sec_Numerical Results}.

\begin{figure}
\begin{centering}
\includegraphics[scale=0.4]{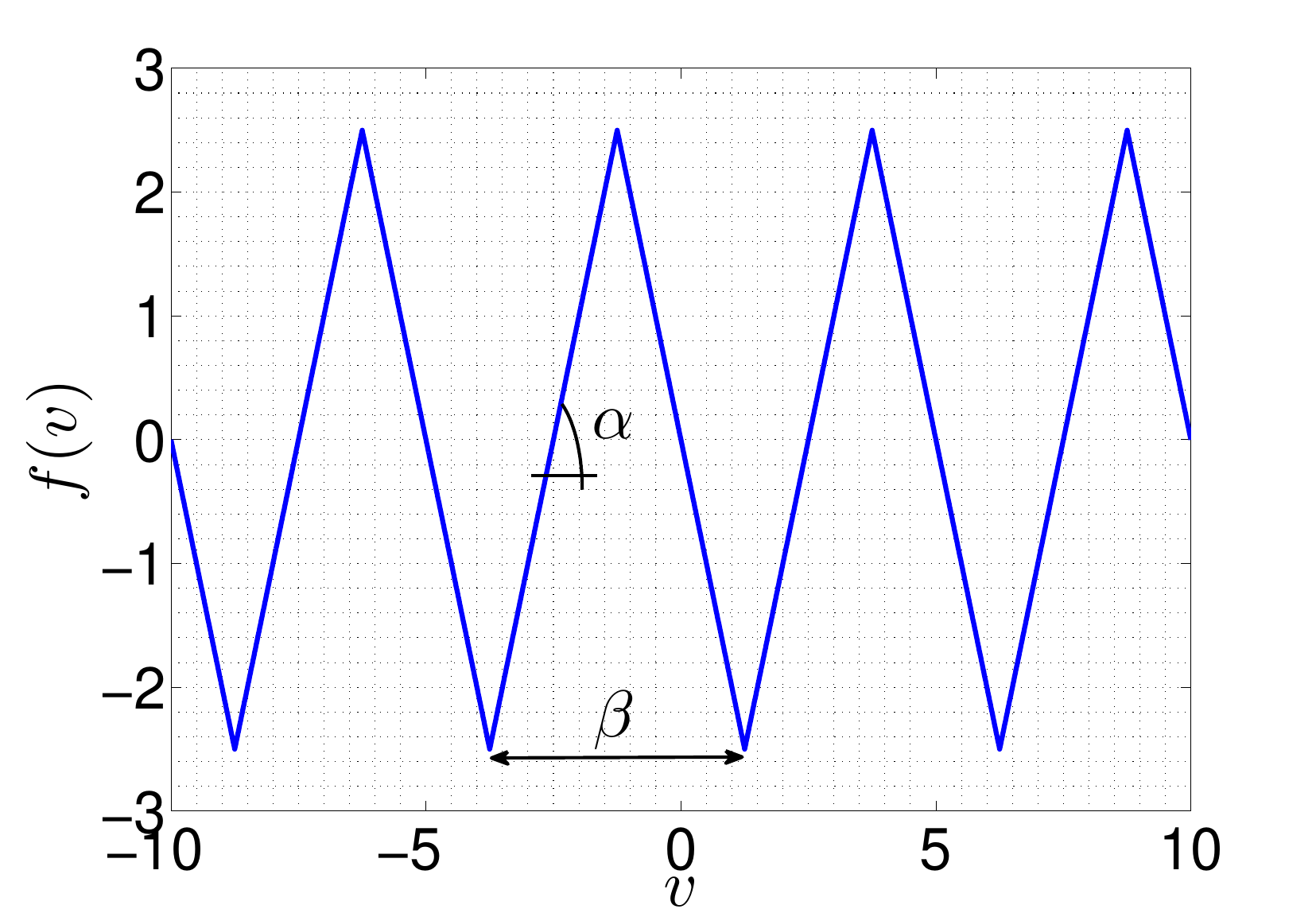}
\caption{Illustration of the PLT encoder mapping for $\alpha=2$ and $\beta=2.5$.}\label{Chptr4:Figure_2}
\par\end{centering}
\vspace{0mm}
\end{figure}

\subsection{Periodic Linear Transmission (PLT)}\label{Chptr4:Sec_linear}
Targeting the low-SNR regime, the first proposed encoder mapping is a periodic linear function with period $\beta$ and slope $\alpha$ within each period. The encoder function is given by
\begin{align}\label{Chptr4:Eqn_7}
f_{\text{PLT}}(v)&=\alpha(-1)^{\left\lfloor\frac{v}{\beta}+\frac{1}{2}\right\rfloor} \left(\beta\left\lfloor\frac{v}{\beta}+\frac{1}{2}\right\rfloor-v\right),
\end{align}
where $\lfloor x\rfloor$ is the largest integer less than or equal to $x$. In Figure \ref{Chptr4:Figure_2}, an illustration of this mapping for $\alpha=2$ and $\beta=2.5$ is shown. To satisfy an average power constraint of $P$, the following condition must be satisfied by $(\alpha,\beta)$
\begin{align}\nonumber
\mathbb{E}[f(V)^2]&=\alpha^2\left(\sigma_v^2+\beta^2\sum\limits_{i=-\infty}^{\infty}i^2\left(Q\left(\frac{-\frac{\beta}{2}+i\beta}{\sigma_v}\right)-Q\left(\frac{\frac{\beta}{2}+i\beta}{\sigma_v}\right)\right)\right.\\
&\left.\quad-\frac{2\beta\sigma_v}{\sqrt{2\pi}} \sum\limits_{i=-\infty}^{\infty} i \left(e^{-\frac{\left(-\frac{\beta}{2}+i\beta\right)^2}{2\sigma_v^2}}-e^{-\frac{\left(\frac{\beta}{2}+i\beta\right)^2}{2\sigma_v^2}}\right)\right)\leq P.
\end{align}
The parameters $\alpha$ and $\beta$ can be optimized under a given average power constraint $P$ in order to minimize the MSE distortion.

\subsection{Periodic BPSK Transmission (PBT)}\label{Chptr4:Sec_digital}
\begin{figure}
\begin{centering}
\includegraphics[scale=0.4]{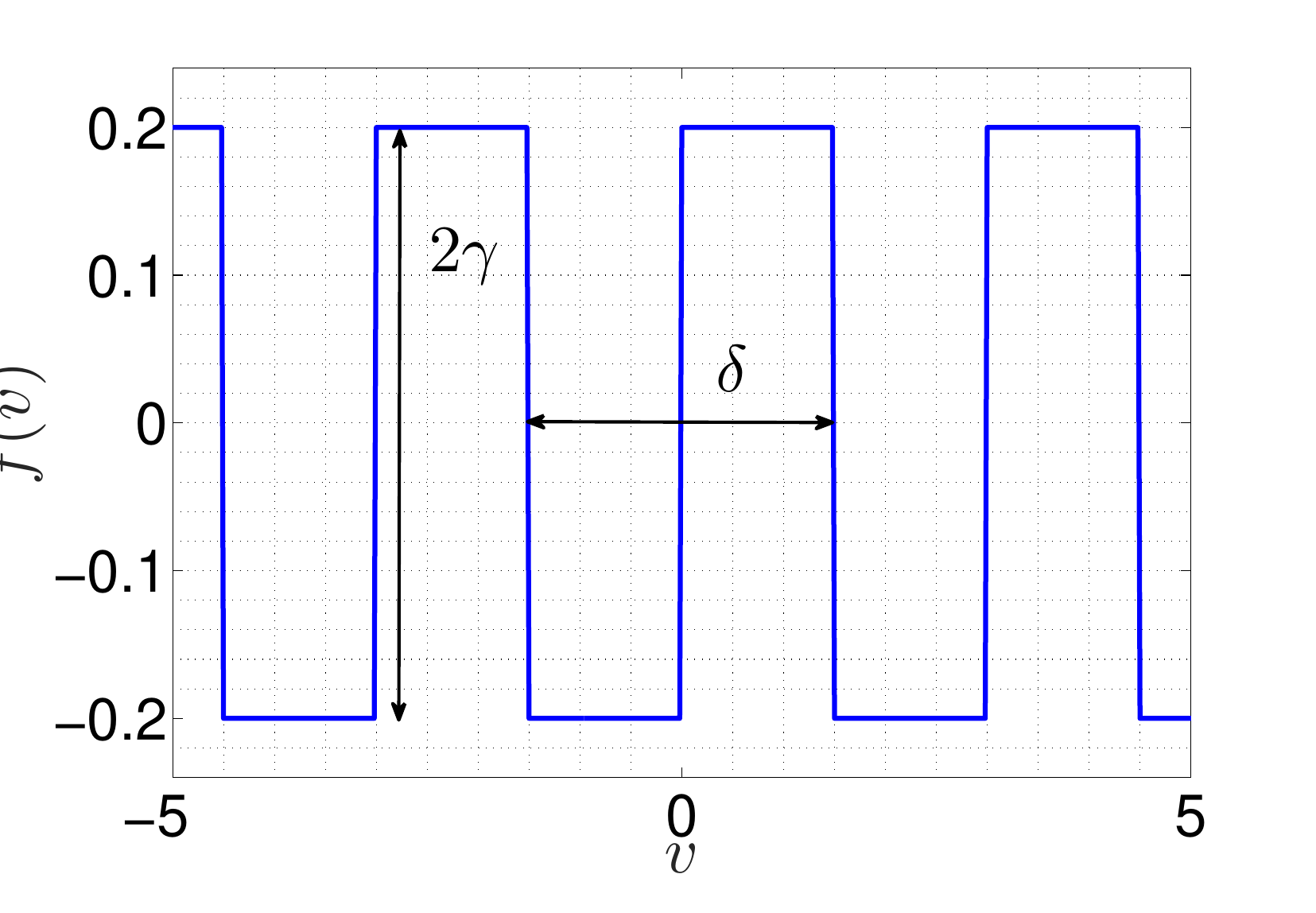}
\caption{Illustration of the PBT encoder mapping for $\gamma=0.2$ and $\delta=3$.}\label{Chptr4:Figure_3}
\par\end{centering}
\vspace{0mm}
\end{figure}

The second proposed encoder mapping, unlike PLT, targets the high-SNR regime and adopts digital modulation with two levels, namely, $\gamma$ and $-\gamma$, with a period of $\delta$. The mapping is defined as
\begin{align}
f_{\text{PBT}}(v)&=\gamma\left(1+2\mathcal{Q}(v)\cdot\text{mod}\left(\left\lfloor\frac{2v}{\delta}\right\rfloor\right)_2\right),
\end{align}
where $\text{mod}(\cdot)_2$ return its argument modulo 2. In Figure \ref{Chptr4:Figure_3}, an illustration of this mapping for $\gamma=0.2$ and $\delta=2.5$ is shown. Due to the average power constraint, we set $\gamma=\sqrt{P}$, and parameter $\delta$ can be optimized to minimize the MSE.

\subsection{Side Information Available at Both the Encoder and Decoder}\label{Chptr4:Sec_preresult}
Here, we consider the scenario in which both the encoder and the decoder have access to the side information $U$. In this case, without loss of optimality, the encoder can encode the error
\begin{equation}\label{Chptr4:Eqn_8}
T=V-\frac{\sigma_v}{\sigma_u} r U,
\end{equation}
where the random variable $\sigma_v r U/\sigma_u$ is the MMSE estimate of $V$ given $U$, which can be computed at both the encoder and the decoder. Since the random variable $T$, which is distributed as $\mathcal{N}(0,\sigma_t^2)$, with $\sigma_t^2=\sigma_v^2(1-r^2)$, is independent of the side information $U$, the encoder can directly encode the error $T$ via a mapping function $\tilde{f}(t)$ ignoring the presence of the side information $U$ at the receiver. Therefore, the problem reduces to the one studied in \cite{arive_varsteh_rassouli_osvaldo_gunduz} and discussed in Section \ref{Chptr4:Sec_Optimal_design} (see Figure \ref{Chptr4:Figure_4}). As a result, a mapping $f(v)=\tilde{f}(v-\sigma_v r u/\sigma_u)$ is optimal, where $\tilde{f}(\cdot)$ is the optimal mapping with no side information shown in Figure \ref{Chptr4:Figure_4}. Therefore, the optimal mapping is centred on the MMSE estimate $\sigma_v r u/\sigma_u$. We will see in Section \ref{Chptr4:Sec_Numerical Results} that, when the side information is not available at the encoder, the NOE consists of periodic replicas of a mapping similar to $\tilde{f}(\cdot)$ in Figure \ref{Chptr4:Figure_4}. As further discussed in Section \ref{Chptr4:Sec_Numerical Results}, the period increases as the variance of the MMSE estimate of $V$ given $U$, namely $\sigma_v^2(1-r^2)$, decreases.

\subsection{Shannon Lower Bound (SLB)}\label{Chptr4:Sec_Lowerbound}
A lower bound on the MSE distortion can be obtained by relaxing the zero-delay constraint, and using the Shannon's source-channel separation theorem. In \cite{Singh_Dabeer_Madhow_2009}, it is shown that the capacity of the AWGN channel with a 1-bit ADC in (\ref{Chptr4:Eqn_1}) is given by 
\begin{align}\label{Chptr4:Eqn_9}
C=1-h\left(Q\left(\sqrt{\text{SNR}}\right)\right),
\end{align}
where $h(\cdot)$ is the binary entropy function defined as $h(p)\triangleq -p\log_2{p}-(1-p)\log_2{(1-p)}$. Furthermore, the rate-distortion function of a Gaussian source with correlated Gaussian side information at the receiver is given by the Wyner-Ziv rate-distortion function \cite{Wyner_Ziv_1976}
\begin{align}\label{Chptr4:Eqn_10}
R(\bar{D})=\frac{1}{2}\left[\log_2{\frac{\sigma_v^2(1-r^2)}{\bar{D}}}\right]^{+},
\end{align}
where $[x]^{+}=\max(0,x)$. Combining (\ref{Chptr4:Eqn_9}) and (\ref{Chptr4:Eqn_10}) a lower bound on the MSE distortion $\bar{D}$ is obtained as
\begin{align}
\bar{D}_{\text{lower}}= (1-r^2)\sigma_v^2 2^{-2\left(1-h\left(Q\left(\sqrt{\text{SNR}}\right)\right)\right)}.
\end{align}

\section{DOP Criterion}\label{Chptr4:Sec_DOP}
In this section, we consider the optimization of the system in Figure \ref{Chptr4_Figure:1} under the DOP criterion. We first derive necessary conditions for an optimal encoder and decoder pair. Then we obtain a lower bound by considering the availability of the side information also at the transmitter.

\subsection{Optimal Encoder and Decoder Design}\label{Chptr4:Sec_Outage_Distortion}
We first obtain the necessary optimality condition of an encoder mapping $f$ for a given decoder $g$. Then, we obtain the optimal decoder $g$ for a given encoder mapping $f$.

\textit{Optimal encoder:} For a fixed decoder function $g(y,u)$, we define the intervals
\begin{align}\label{Chptr4:Eqn_11}
I_y(u)=\left\{v:(v-g(y,u))^2< D\right\},~ y=0,1.
\end{align}
Each interval $I_0(u)$ and $I_1(u)$ in (\ref{Chptr4:Eqn_11}) corresponds to the set of source values that are within the allowed distortion target $D$ of the reconstruction points $g(0,u)$ and $g(1,u)$, respectively, when the side information is $U=u$. Hence, the following claims hold: (\textit{i}) For all source realizations $v$ in the set $(I_0(u)\cup I_1(u))^{C}=\{v:\min_{y=0,1}(v-g(y,u))^2 \geq D\}$, outage occurs since no reconstruction point $g(y,u)$ satisfies the distortion constraint (superscript $C$ denotes the complement set). We refer to this event as \textit{source outage}. (\textit{ii}) For all source realizations in the interval $I_0(u) \cap I_1(u)$, either of the reconstruction points yields a distortion not larger than the target value $D$. Therefore, regardless of which output $(g(0,u),g(1,u))$ is selected by the receiver, no outage occurs.

With these observations in mind, the next proposition characterizes the optimal encoder mapping $f$ for a given decoder $g$.

\begin{prop}\label{Chptr4:Prop_2}
Given a target distortion $D$, and a decoder with reconstruction function $g(\cdot,\cdot)$, the optimal mapping $f(\cdot)$ for the problem (\ref{Chptr4:Eqn_3}) under the DOP criterion satisfies
\begin{align}\label{Chptr4:Eqn_12}
f(v)&=\frac{e^{-\frac{f(v)^2}{2\sigma_w^2}}}{2\lambda\sqrt{2\pi}}\Big(\textrm{Pr}\left(U\in S_{0\setminus1}(v)\right)-\textrm{Pr}\left(U\in S_{1\setminus0}(v)\right)\Big).
\end{align}
where $S_{0\setminus1}(v)$ and $S_{1\setminus 0}(v)$ are defined as
\begin{align}\nonumber
S_{0\setminus1}(v)&\triangleq\{u:~b_{0l}(u)\leq v\leq b_{0r}(u) \},\\\label{Chptr4:Eqn_13}
S_{1\setminus0}(v)&\triangleq\{u:~b_{1l}(u)\leq v\leq b_{1r}(u)\},
\end{align}
and $b_{1r}(u)$, $b_{1l}(u)$, $b_{0r}(u)$ and $b_{0l}(u)$ are defined as below
\begin{subequations}\label{Chptr4:Eqn_14}
\begin{align}
b_{0r}(u)&\triangleq\left\{\begin{array}{ll}
             g(0,u)+\sqrt{D} & g(0,u)\geq g(1,u),\\
             \min\left\{g(1,u)-\sqrt{D},g(0,u)+\sqrt{D}\right\} & g(0,u)< g(1,u),
           \end{array}\right.\\
b_{0l}(u)&\triangleq\left\{\begin{array}{ll}
             \max\left\{g(1,u)+\sqrt{D},g(0,u)-\sqrt{D}\right\} & g(0,u)\geq g(1,u),\\
             g(0,u)-\sqrt{D} & g(0,u)< g(1,u),
           \end{array}\right.\\
b_{1r}(u)&\triangleq\left\{\begin{array}{ll}
             \min\left\{g(1,u)+\sqrt{D},g(0,u)-\sqrt{D}\right\} & g(0,u)\geq g(1,u),\\
             g(1,u)+\sqrt{D} & g(0,u)< g(1,u),
           \end{array}\right.\\
b_{1l}(u)&\triangleq\left\{\begin{array}{ll}
             g(1,u)-\sqrt{D} & g(0,u)\geq g(1,u),\\
             \max\left\{g(1,u)-\sqrt{D},g(0,u)+\sqrt{D}\right\} & g(0,u)< g(1,u).
           \end{array}\right.
\end{align}
\end{subequations}
Furthermore, the gradient of the Lagrangian function $L(f,g,\lambda)$ over $f$, for a given $g$, is found as
\begin{align}\label{Chptr4:Eqn_40}
\nabla L&=2\lambda f(v)-\frac{e^{-\frac{f(v)^2}{2\sigma_w^2}}}{\sqrt{2\pi}}\Big(\textrm{Pr}\left(U\in S_{0\setminus1}(v)\right)-\textrm{Pr}\left(U\in S_{1\setminus0}(v)\right)\Big).
\end{align}

\end{prop}
\textit{Proof}: See Appendix \ref{Chptr4:Appendix_4}.

\textit{Optimal decoder:} Assuming that the encoder mapping $f$ is given, we now aim to minimize the Lagrangian function in (\ref{Chptr4:Eqn_3}) for the DOP criterion over the decoding function $g$. The next proposition characterizes the optimal decoder mapping for a given encoder $f$.

\begin{prop}\label{Chptr4:Prop_3}
Given a target distortion $D$ and an encoder mapping $f(\cdot)$, the optimal decoder $g(\cdot,\cdot)$ for the problem (\ref{Chptr4:Eqn_3}) under the DOP criterion is obtained as
\begin{align}\label{Chptr4:Eqn_15}
g(y,u)&\in~~\stackunder[5pt]{arg max}{$\hat{v}$}~~\int\limits_{\hat{v}-\sqrt{D}}^{\hat{v}+\sqrt{D}}\Phi\left(\frac{v}{\sigma_v}\Big|\frac{u}{\sigma_u}\right)Q\left(\frac{(-1)^{y+1}f(v)}{\sigma_w}\right)dv.
\end{align}
\end{prop}
\textit{Proof}: See Appendix \ref{Chptr4:Appendix_5}.

To provide further insights into Propositions \ref{Chptr4:Prop_2} and \ref{Chptr4:Prop_3}, it is worth considering the two extreme cases of side information correlation. When $r=0$ the decoder outputs $g(0,u)=\hat{v}_0$ and $g(1,u)=\hat{v}_1$ are independent of the side information $U$, and the conditions derived here coincide with those obtained in \cite[Proposition IV.1]{arive_varsteh_rassouli_osvaldo_gunduz} for the optimal mapping when there is no side information at the receiver. Instead, with $r=1$, we can choose $g(y,u)=g(u)=\pm \frac{\sigma_v}{\sigma_u}u$; and hence, we have $f(v)=0$ for all $v$.

\begin{rem}\label{Chptr4:Remark_4}
In the low SNR regime, from (\ref{Chptr4:Eqn_15}), we have $g(y,u)\simeq\frac{r\sigma_v}{\sigma_u}u$, $y=0,1$. Therefore, in the asymptotic low SNR regime, the DOP at the receiver is found as (see Appendix \ref{Chptr4:Appendix_6})
\begin{align}\label{Chptr4:Eqn_41}
 \lim_{\text{SNR}\rightarrow 0} \epsilon(D)=2Q\left(\frac{\sqrt{D}}{\sigma_v\sqrt{1-r^2}}\right).
\end{align}
In Section \ref{Chptr4:Sec_Numerical Results}, we validate (\ref{Chptr4:Eqn_41}) in the asymptotic low SNR regime.
\end{rem}

\subsection{Side Information Available at Both Encoder and Decoder}\label{Chptr4:Sec_Previous Result}
When the side information $U$ is also available at the encoder, using the optimal decoder under the DOP criterion, the encoder can reconstruct the source as
\begin{subequations}\label{Chptr4:Eqn_43}
\begin{align}
&\arg \min_{\hat{v}} \textrm{Pr}(|V-\hat{v}|^2\geq D|U=u)\\
&\quad\quad=\arg \max\limits_{\hat{v}} \int\limits_{\hat{v}-\sqrt{D}}^{\hat{v}+\sqrt{D}}\Phi\left(\frac{v}{\sigma_v}\Big|\frac{u}{\sigma_u}\right)dv\\\label{Chptr4:Eqn_44}
&\quad\quad=\frac{\sigma_v}{\sigma_u}ru.
\end{align}
\end{subequations}
From (\ref{Chptr4:Eqn_43}), we conclude that the best estimate of a Gaussian source from Gaussian side information under the DOP criterion equals the MMSE estimate. Given that (\ref{Chptr4:Eqn_44}) can be reconstructed at both encoder and decoder, as in Section \ref{Chptr4:Sec_preresult}, the optimal encoder uses the optimal mapping for the scenario without side information \cite{arive_varsteh_rassouli_osvaldo_gunduz} as applied to the error signal in (\ref{Chptr4:Eqn_8}). In the following proposition, we show that the optimal decoder under the DOP criterion is obtained by summing the estimates computed on the basis of the side information $u$ and the channel output $y$, separately.

\begin{prop}\label{Chptr4:Prop_4}
Given a target distortion $D$, the optimal decoder $g(\cdot,\cdot)$ for the problem (\ref{Chptr4:Eqn_3}) under the DOP criterion is obtained as
\begin{align}
g(y,u)&=\frac{r\sigma_v}{\sigma_u}u+\hat{t}_y,~y=0,1,
\end{align}
where $\hat{t}_y$ represents the optimal decoder for a Gaussian source with variance $(1-r^2)\sigma_v^2$ as a function of $Y$, which is given in \cite[Proposition IV.2]{arive_varsteh_rassouli_osvaldo_gunduz}. The resulting minimum DOP is obtained as
\begin{align}
\epsilon(D)&=2Q\left(\frac{2\sqrt{D}-a}{\sqrt{1-r^2}\sigma_v}\right)+2Q\left(\frac{t}{\sigma_w}\right)
\left(Q\left(\frac{a}{\sqrt{1-r^2}\sigma_v}\right)-Q\left(\frac{2\sqrt{D}-a}{\sqrt{1-r^2}\sigma_v}\right)\right),
\end{align}
where $t$ is the solution of the equation $t e^{\frac{t^2}{2\sigma_w^2}}=\frac{1}{2\sqrt{2\pi}\sigma_w\lambda}$.
\end{prop}
\textit{Proof}: See Appendix \ref{Chptr4:Appendix_7}.

\begin{figure}
\begin{centering}
\includegraphics[scale=0.4]{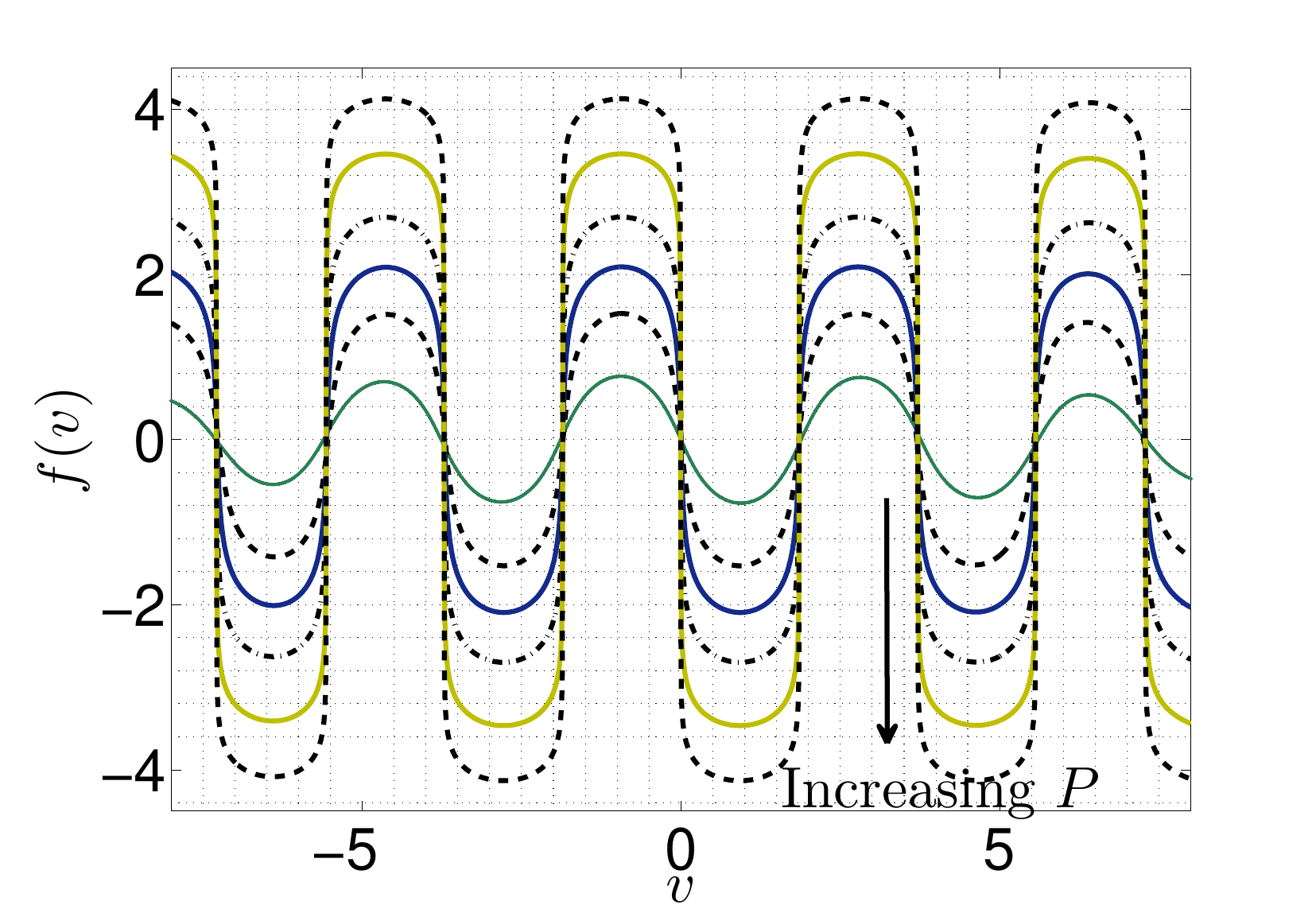}
\caption{NOE mappings under the MSE criterion with different average power values
and $r = 0.85$ $(\sigma_v = \sigma_w = 1)$. Increasing the power constraint $P$ has no
impact on the period of the NOE mapping, which instead depends on $r$ (see Figure \ref{Chptr4:Figure_7}).}\label{Chptr4:Figure_5}
\par\end{centering}
\vspace{0mm}
\end{figure}

\section{Numerical Results}\label{Chptr4:Sec_Numerical Results}

In this section, we present numerical results with the aim of assessing the performance of the encoder/ decoder pairs obtained in the previous sections. In order to derive the NOE mappings we apply a gradient descent-based iterative algorithm. The algorithm performs a gradient descent search in the opposite direction of the derivative of the Lagrangian (\ref{Chptr4:Eqn_3}) with respect to the encoder mapping $f(\cdot)$. The update is obtained by
\begin{align}
  f_{i+1}(v)=f_i(v)-\mu \nabla_f L,
\end{align}
where $i$ is the iteration index, $\nabla_f L$ is defined in (\ref{Chptr4:Eqn_39}) and (\ref{Chptr4:Eqn_40}) for the MSE distortion and DOP criterion, respectively, and $\mu>0$ is the step size. The algorithm can be initialized with an arbitrary mapping. Here, we use a linear mapping with slope close to zero for initialization. It is noted that the algorithm is not guaranteed to converge to a global optimal solution. We also remark that different power constraints are imposed by means of a linear search over the Lagrange multiplier $\lambda$. In the following, we first discuss the numerical results for the MSE distortion criterion, followed by the DOP criterion.

\begin{figure}
\begin{centering}
\includegraphics[scale=0.4]{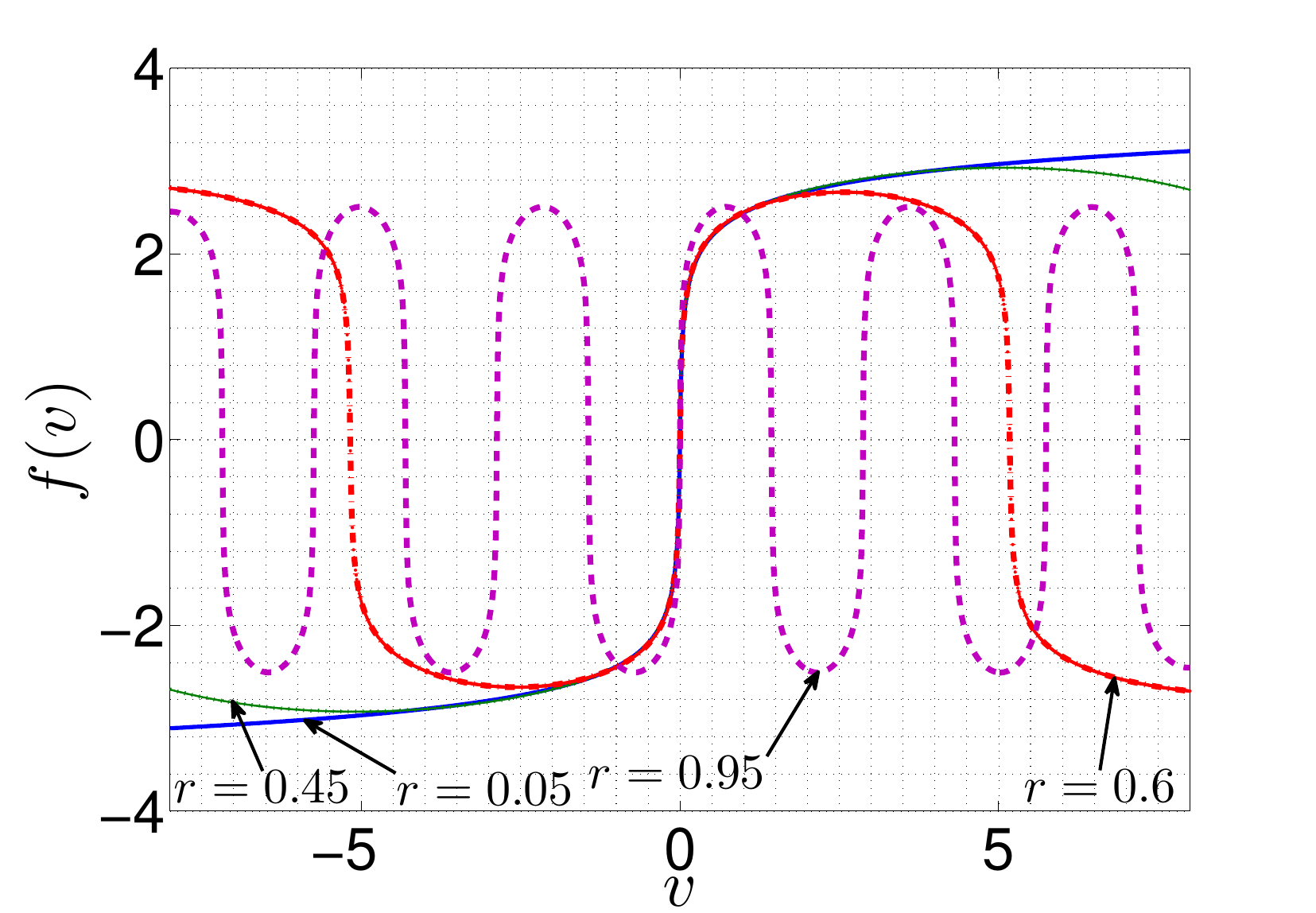}
\caption{NOE mappings under the MSE criterion for different
correlation coefficients $r$ and an average power constraint $P = 5$
$(\sigma_v^2 = \sigma_w ^2= 1).$}\label{Chptr4:Figure_7}
\par\end{centering}
\vspace{0mm}
\end{figure}

\begin{figure}
\begin{centering}
\includegraphics[scale=0.6]{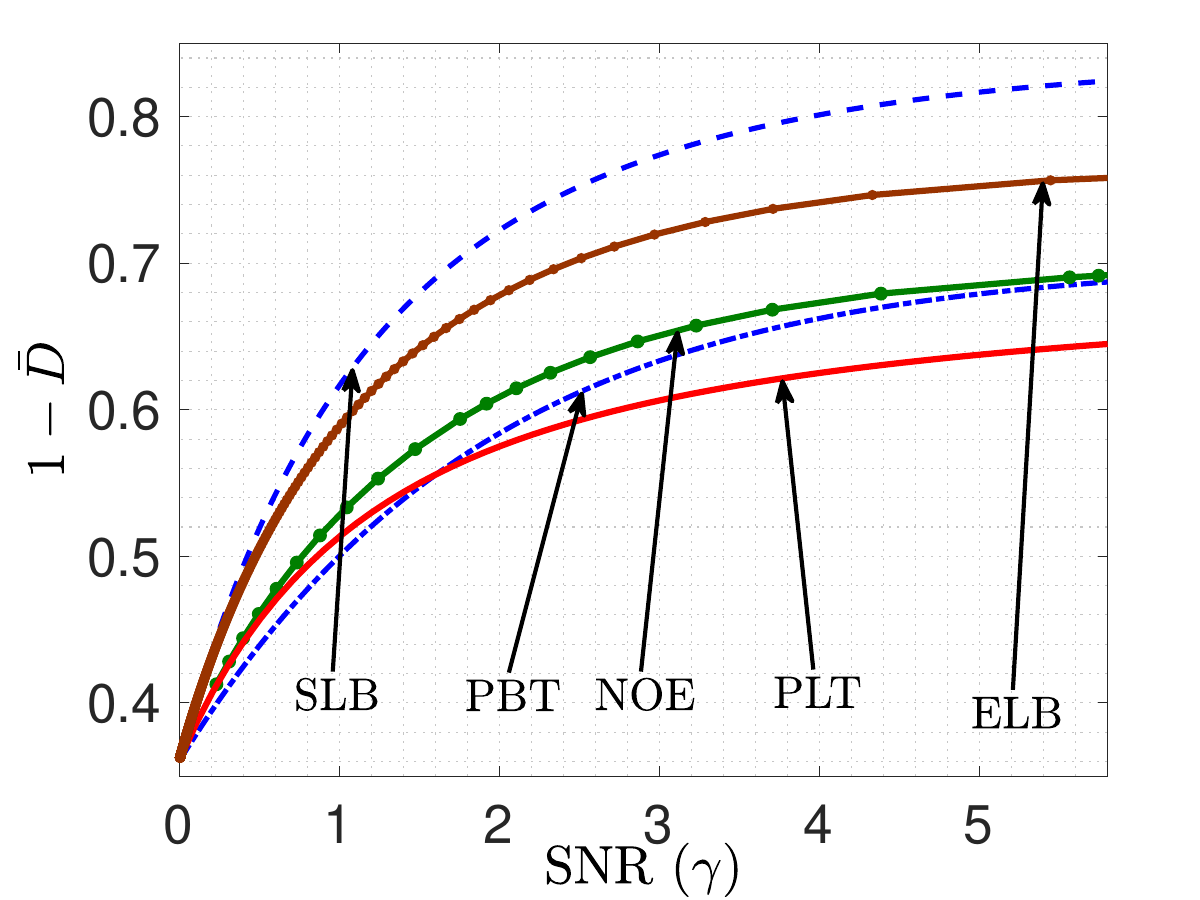}
\caption{Complementary MSE distortion vs. SNR for $r = 0.6~(\sigma_v^2 = \sigma_w ^2=
1)$.}\label{Chptr4:Figure_6}
\par\end{centering}
\vspace{0mm}
\end{figure}

\textit{MSE criterion}: In Figure \ref{Chptr4:Figure_5}, NOE mappings for the MSE distortion criterion obtained using the aforementioned gradient descent algorithm are plotted for different average power constraints, for a correlation coefficient of $r=0.85$. We note the periodic structure of the mappings, which is in line with the results in \cite{Akyol_Viswanatha_Rose_2014} for an infinite resolution front end. In contrast, the optimal mapping obtained in \cite{arive_varsteh_rassouli_osvaldo_gunduz} when $r=0$ is a monotonically increasing function (see Figure \ref{Chptr4:Figure_4}). We also observe that the average power constraint $P$ does not affect the period of the mapping. In Figure \ref{Chptr4:Figure_7}, NOE mappings for an average power of $P=5$ are plotted for different correlation coefficients. We see that the period of the mapping instead depends on $r$: the higher the correlation coefficient $r$ is, the smaller the period of the mapping is. Furthermore, Figure \ref{Chptr4:Figure_5} shows that the SNR, or $P$, affects the slope of the encoder mapping in each period in a manner similar to Figure \ref{Chptr4:Figure_4}, so that, for low SNR the optimal mapping resembles the perioding linear mappings studied in Section \ref{Chptr4:Sec_linear}, while for high SNR, the optimal mapping resembles the periodic BPSK mappings studied in Section \ref{Chptr4:Sec_digital}.

\begin{figure}
\begin{centering}
\includegraphics[scale=0.6]{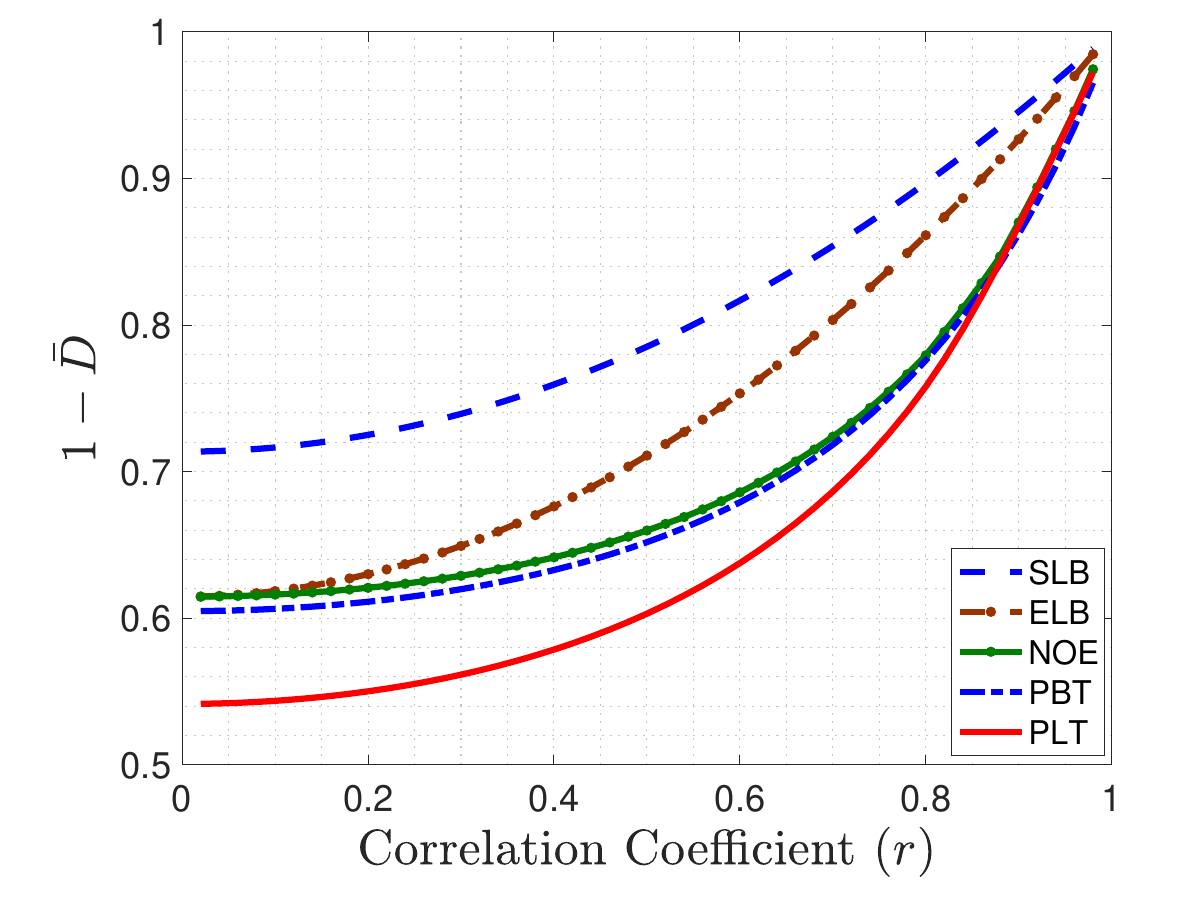}
\caption{Complementary MSE distortion versus correlation coefficient $r$ under
the average power constraint $P = 5~(\sigma_v^2 = \sigma_w ^2=1)$.}\label{Chptr4:Figure_8}
\par\end{centering}
\vspace{0mm}
\end{figure}

In Figure \ref{Chptr4:Figure_6}, we plot the complementary MSE distortion $(1-\bar{D})$ versus SNR for the NOE,  as well as for the PLT and PBT schemes, for correlation coefficient $r=0.6$. The SLB and the MSE distortion achieved when both the encoder and the decoder have access to the side information $U$, which is referred to as the encoder lower bound (ELB), are also included for comparison. Following the discussion above, we observe that the performance of PBT is close to that of NOE at high SNR values. On the other hand, for low SNRs, PLT outperforms PBT and approaches the NOE performance. In Figure \ref{Chptr4:Figure_8}, we plot the complementary MSE distortion $(1-\bar{D})$ versus the correlation coefficient $r$ for a fixed average power constraint of $P=5$. We note from Figure \ref{Chptr4:Figure_8} that the ELB is tight in the low and high correlation regime, and in general there is a loss in the MSE distortion by not having the side information at the encoder. We recall that this is not the case with infinite resolution and infinite block-length. We also observe that PLT is tighter when the correlation is higher, while it performs quite poorly when the side information quality is poor. On the other hand, for this $P$ values PBT performs relatively close to NOE for the whole range of side information correlation values.

\begin{figure}
\begin{centering}
\includegraphics[scale=0.4]{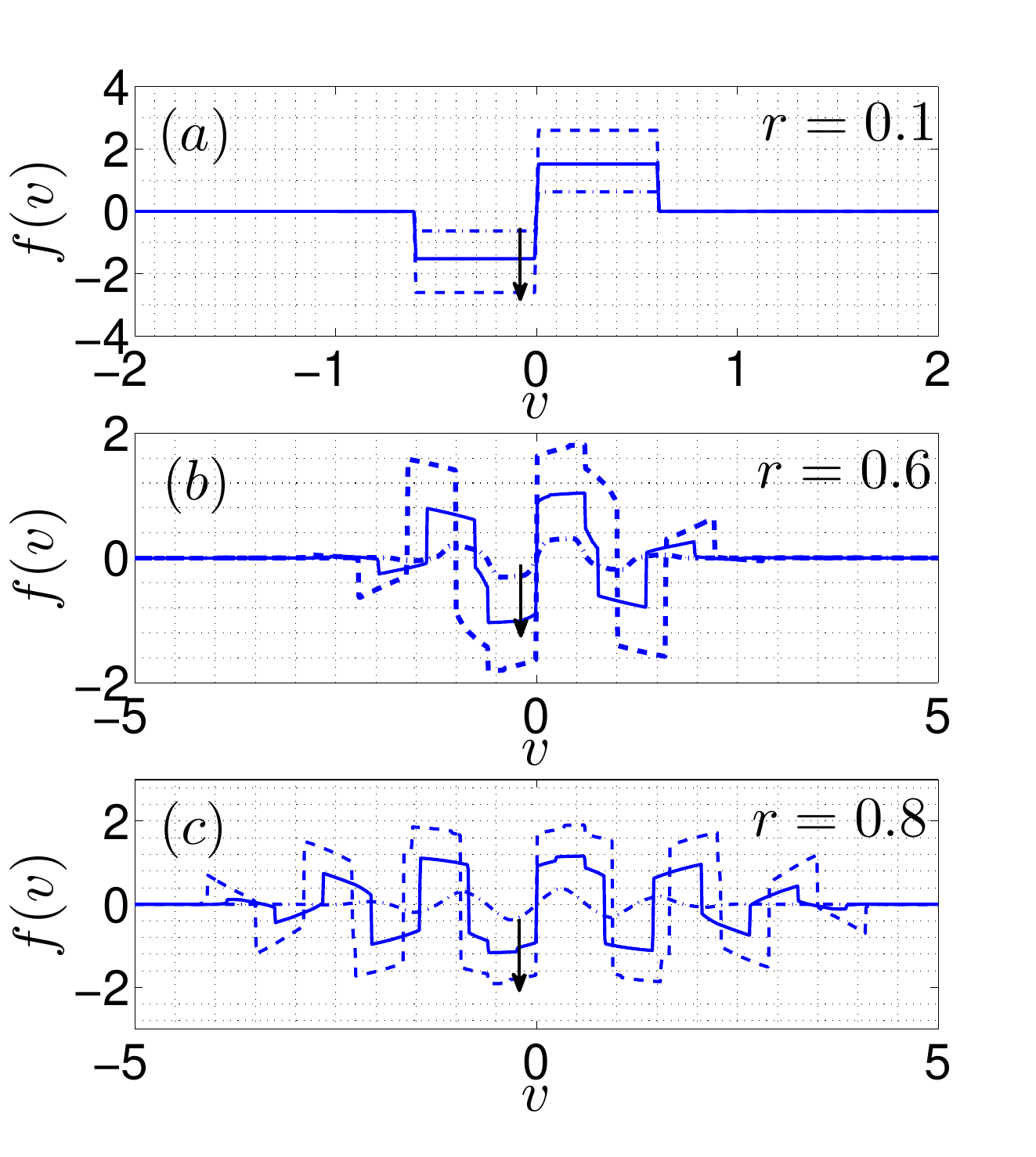}
\caption{NOE mappings under the DOP criterion for different correlation coefficients, $(\sigma_v^2 = \sigma_w ^2=1)$. The power constraint $P$ of the mappings increases in the direction of the arrow.}\label{Chptr4:Figure_9}
\par\end{centering}
\vspace{0mm}
\end{figure}

\textit{DOP criterion}: In Figure \ref{Chptr4:Figure_9}, NOE mappings for different power constraints and correlation coefficients are shown under the DOP criterion. For low enough values of the correlation coefficient, such as $r=0.1$, the NOE mappings resemble the optimal mappings in the absence of receiver side information obtained in \cite{arive_varsteh_rassouli_osvaldo_gunduz}, which corresponds to a binary transmitter as seen in Figure \ref{Chptr4:Figure_9}. We observe that the domain of the mapping is limited, unlike for the MSE criterion, since values of the source that differ by more than $\sqrt{D}$ from the reconstruction points yield an outage irrespective of the mapping. As the correlation between the source and the side information increases, the domain of the mapping expands.

\begin{figure}
\begin{centering}
\includegraphics[scale=0.6]{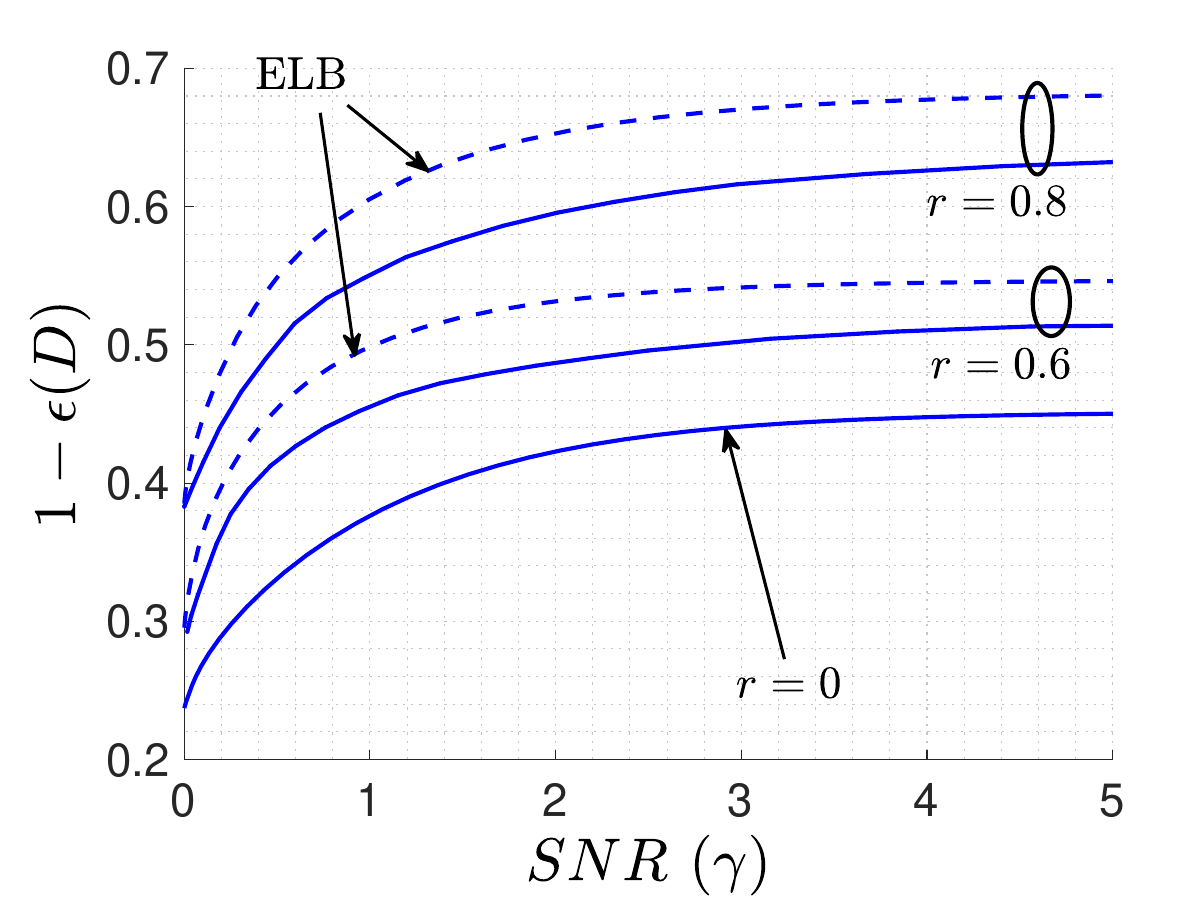}
\caption{Complementary DOP vs. SNR for $r =0,~0.6,~0.8 ~(\sigma_v^2 = \sigma_w^2 =
1)$ and $D=0.09$.}\label{Chptr4:Figure_10}
\par\end{centering}
\vspace{0mm}
\end{figure}

In Figure \ref{Chptr4:Figure_10}, we plot the complementary DOP, $1-\epsilon(D)$, versus SNR for NOE mappings as well as the ELB under the DOP criterion, for correlation coefficients $r=0,~0.6,~0.8$. We observe that in the low SNR regime the DOP is close to the ELB. This is because, in the low SNR regime the source estimate can be obtained  based mainly on the side information. We also observe that, as the SNR increases, the DOP saturates to the source outage probability, which is independent of the SNR.

\section{Conclusions}\label{Chptr4:Sec_conclusion}
We considered the problem of transmission of a Gaussian source over an AWGN channel to a receiver equipped with a one-bit ADC front end. We also considered the availability of a correlated side information at the receiver. We studied this problem under two distinct performance criteria, namely the MSE distortion and the DOP, while imposing an average power constraint at the transmitter. Assuming that the transmission is zero-delay, in the sense that, it maps every single source output to a single channel input, we obtained necessary conditions for the optimal encoder and decoder mappings under both performance criteria. In the comparison to the previous work in \cite{arive_varsteh_rassouli_osvaldo_gunduz}, we observed that the availability of correlated side information at the receiver has a significant impact on the shape of the optimal encoder mapping. For instance, as in the case of infinite-resolution front end \cite{Akyol_Viswanatha_Rose_2014}, the optimal mapping becomes periodic under the MSE distortion criterion. We observed that the period of the optimal mapping depends on the correlation coefficient between the source and the side information, but it is not affected by the transmitter power condition. For the DOP criterion, the availability of the side information enlarges the domain of the mapping, i.e., a larger set of source sample values are mapped to a non-zero channel input. Interesting future research direction include investigating the effect of higher level ADCs on the performance of the system, obtaining optimized structures when there is fading in the channel or when there are multi observations at the receiver.

\section{Appendices}
\appendix 
\section{Preliminaries: Calculus of Variations}
In the proofs of the Propositions \ref{Chptr4:Prop_1} and \ref{Chptr4:Prop_2}, we use variational calculus \cite[Section 7]{Luenberger_1969} to obtain necessary optimality conditions. The next theorem summarizes the key result that will be needed.

\begin{thm}
Let $F$, $G_0$ and $G_i,~i=1,\ldots,n$, be continuous functionals of $(f,H,t)$, $(f,r_1,\ldots,r_n,u,t)$ and $(f,t,u)$, respectively, where $H$ and $r_i,~i=1,\ldots,n$, are given by
\begin{align}
H(t)&=\int\limits_{t_1}^{t_2} G_0(f(t),r_1(u),\ldots, r_n(u),u,t)du,\\
r_i(u)&=\int\limits_{t_1}^{t_2} G_i(f(v),v,u)dv,~~i=1,\ldots,n.
\end{align}
Also, let $F$, $G_0$ and $G_i,~i=1,\ldots,n$, have continuous partial derivatives with respect to $(f,H)$, $(f,r_1,\ldots,r_n)$ and $f$, respectively. Consider the following minimization problem
\begin{eqnarray}\label{Chptr4:Eqn_18}
\begin{aligned}
& \underset{f}{\text{minimize}}
& & L(f)\triangleq\int\limits_{t_1}^{t_2} F(f(t),H(t),t)dt.
\end{aligned}
\end{eqnarray}
Define $\nabla L$ as
\begin{align}\nonumber
\nabla L&\triangleq F^{f}(f(t),H(t),t)+F^{H}(f(t),H(t),t)\int\limits_{t_1}^{t_2} G_0^{f}(f(t),r_1(u),\ldots, r_n(u),u,t)du\\\label{Chptr4:Eqn_19}
&\quad +\int\limits_{t_1}^{t_2}\int\limits_{t_1}^{t_2}F^{H}(f(v),H(v),v)\sum\limits_{i=1}^{n}G_i^{f}(f(t),t,u) G_0^{r_i}(f(v),r_1(u),\ldots, r_n(u),u,v)dvdu,
\end{align}
where $F^f$and $F^H$ denote the partial derivatives of the functional $F$ with respect to $f$ and $H$, respectively; and $G_0^f$ and $G_0^{r_i}$ denote the partial derivatives of the functional $G_0$  with respect to $f$ and $r_i$, respectively. Similarly, $G_i^f$ denotes the partial derivative of the functional $G_i$ with respect to $f$. A necessary condition for a function $f$ to be a solution to the minimization problem in (\ref{Chptr4:Eqn_18}) is
\begin{align}
  \nabla L=0.
\end{align}
\end{thm}

\textit{Proof}:
Following the conventional approach in the calculus of variations, we perturb the function $f(t)$ by an arbitrary function $\eta(t)$ that vanishes on the boundary points $t_1$ and $t_2$ \cite{Luenberger_1969}. Let $\delta_f L\triangleq \frac{dL(f+\alpha\eta)}{d\alpha}\Big|_{\alpha=0}$ be the resulting Gateaux derivative of the functional $L$ with respect to the parameter $\alpha$. We have
\begin{align}\label{Chptr4:Eqn_20}
\delta_f L&=\frac{d}{d\alpha}\int\limits_{t_1}^{t_2} F(f(t)+\alpha \eta(t),H^\alpha(t) ,t)dt\Bigg|_{\alpha=0},
\end{align}
where $H^\alpha(t)$ is defined as
\begin{align}
H^\alpha(t)&\triangleq\int\limits_{t_1}^{t_2} G_0(f(t)+\alpha\eta(t),r_1^\alpha(u),\ldots, r_n^\alpha(u),u,t)du,
\end{align}
and $r_i^\alpha(u),~i=1,\ldots,n$, are defined as
\begin{align}
r_i^\alpha(u)&\triangleq\int\limits_{t_1}^{t_2} G_i(f(v)+\alpha\eta(v),v,u)dv,~~i=1,\ldots,n.
\end{align}
Evaluating the derivative in (\ref{Chptr4:Eqn_20}), we have
\begin{align}\label{Chptr4:Eqn_21}
\delta_f L&=\int\limits_{t_1}^{t_2} \left[\eta(t) F^{f}(f(t),H(t),t)+\frac{dH^\alpha(t)}{d\alpha}F^{H}(f(t),H(t),t)\right]dt,
\end{align}
where $\frac{dH^\alpha(t)}{d\alpha}$ is the Gateaux derivative of the functional $H(t)$, which is computed as
\begin{align}
\frac{dH^\alpha(t)}{d\alpha}&=\frac{d}{d\alpha}\int\limits_{t_1}^{t_2} G_0(f(t)+\alpha\eta(t),r_1^\alpha(u),\ldots, r_n^{\alpha}(u),u,t)du\Big|_{\alpha=0}\\\nonumber
&=\int\limits_{t_1}^{t_2}\left(\eta(t) G_0^{f}(f(t),r_1(u),\ldots, r_n(u),u,t)\right.\\\label{Chptr4:Eqn_22}
&\left.\quad+ \sum\limits_{i=1}^{n} \frac{dr_i^{\alpha}(u)}{d\alpha}  G_0^{r_i}(f(t),r_1(u),\ldots, r_n(u),u,t)\right)du,
\end{align}
where
\begin{align}
\frac{dr_i^{\alpha}(u)}{d\alpha}&\triangleq\frac{d}{d\alpha}\int\limits_{t_1}^{t_2} G_i(f(v)+\alpha\eta(v),v,u)dv\Bigg|_{\alpha=0}\\\label{Chptr4:Eqn_23}
&=\int\limits_{t_1}^{t_2} \eta(v) G_i^{f}(f(v),v,u)dv.
\end{align}
By plugging (\ref{Chptr4:Eqn_23}) into (\ref{Chptr4:Eqn_22}), we can write
\begin{align}\nonumber
\frac{dH^\alpha(t)}{d\alpha}&=\int\limits_{t_1}^{t_2}\eta(t) G_0^{f}(f(t),r_1(u),\ldots, r_n(u),u,t)du,\\\label{Chptr4:Eqn_24}
&\quad+\int\limits_{t_1}^{t_2}\int\limits_{t_1}^{t_2}\sum\limits_{i=1}^{n}  \eta(v)G_i^{f}(f(v),v,u) G_0^{r_i}(f(t),r_1(u),\ldots, r_n(u),u,t)dvdu.
\end{align}

By substituting (\ref{Chptr4:Eqn_24}) into (\ref{Chptr4:Eqn_21}) we have
\begin{align}\nonumber
\delta_f L&=\int\limits_{t_1}^{t_2} \eta(t)F^{f}(f(t),H(t),t)dt\\\nonumber
&\quad+\int\limits_{t_1}^{t_2}F^{H}(f(t),H(t),t)\left(\int\limits_{t_1}^{t_2}\eta(t) G_0^{f}(f(t),r_1(u),\ldots, r_n(u),u,t)du\right.\\
&\quad\left.+\int\limits_{t_1}^{t_2}\int\limits_{t_1}^{t_2}\sum\limits_{i=1}^{n}  \eta(v)G_i^{f}(f(v),v,u) G_0^{r_i}(f(t),r_1(u),\ldots, r_n(u),u,t)dvdu\right)dt\\\nonumber
&=\int\limits_{t_1}^{t_2} \eta(t)\left(F^{f}(f(t),H(t),t)+F^{H}(f(t),H(t),t)\int\limits_{t_1}^{t_2} G_0^{f}(f(t),r_1(u),\ldots, r_n(u),u,t)du\right)dt\\
&\quad+\int\limits_{t_1}^{t_2}\int\limits_{t_1}^{t_2}\int\limits_{t_1}^{t_2}\eta(v)F^{H}(f(t),H(t),t)\sum\limits_{i=1}^{n}G_i^{f}(f(v),v,u) G_0^{r_i}(f(t),r_1(u),\ldots, r_n(u),u,t)dvdudt\\\nonumber
&=\int\limits_{t_1}^{t_2} \eta(t)\left(F^{f}(f(t),H(t),t)+F^{H}(f(t),H(t),t)\int\limits_{t_1}^{t_2} G_0^{f}(f(t),r_1(u),\ldots, r_n(u),u,t)du\right)dt\\
&\quad+\int\limits_{t_1}^{t_2}\int\limits_{t_1}^{t_2}\int\limits_{t_1}^{t_2}\eta(t)F^{H}(f(v),H(v),v)\sum\limits_{i=1}^{n}G_i^{f}(f(t),t,u) G_0^{r_i}(f(v),r_1(u),\ldots, r_n(u),u,v)dvdudt\\\nonumber
&=\int\limits_{t_1}^{t_2} \eta(t)\left(F^{f}(f(t),H(t),t)+F^{H}(f(t),H(t),t)\int\limits_{t_1}^{t_2} G_0^{f}(f(t),r_1(u),\ldots, r_n(u),u,t)du\right.\\\label{Chptr4:Eqn_25}
&\left.\quad+\int\limits_{t_1}^{t_2}\int\limits_{t_1}^{t_2}F^{H}(f(v),H(v),v)\sum\limits_{i=1}^{n}G_i^{f}(f(t),t,u) G_0^{r_i}(f(v),r_1(u),\ldots, r_n(u),u,v)dvdu\right)dt.
\end{align}
Since $\eta(t)$ in (\ref{Chptr4:Eqn_25}) is an arbitrary function, the necessary condition for $f$ to be a solution is that the term inside the round brackets in (\ref{Chptr4:Eqn_25}) is zero. This concludes the proof.
\qed

\section{Proof of Proposition \ref{Chptr4:Prop_1} }\label{Chptr4:Appendix_2}
Due to the orthogonality principle of the MMSE estimation, it can be easily verified that $\bar{D}=\sigma_v^2-\mathbb{E}[V\hat{V}]$. Rewriting the Lagrangian $L(f,g,\lambda)$ for the MSE distortion criterion and dropping constants that are independent of $f$, we have
\begin{eqnarray}\label{Chptr4:Eqn_26}
\stackunder[5pt]{minimize}{$f$}~-\mathbb{E}[V\hat{V}]+\lambda\mathbb{E}[f(V)^2].
\end{eqnarray}
By expanding the objective function in (\ref{Chptr4:Eqn_26}), it can be written as
\begin{align}\nonumber
&\frac{-1}{\sigma_w\sigma_v\sigma_u}\int \int\int v g(y,u)\Phi\left(\frac{v}{\sigma_v},\frac{u}{\sigma_u}\right)\Phi\left(\frac{w}{\sigma_w}\right)dwdudv\\
&\quad+\frac{\lambda}{\sigma_v}\int\Phi\left(\frac{v}{\sigma_v}\right) f(v)^2 dv\\\nonumber
&=\frac{-1}{\sigma_v\sigma_u}\int\int v\left(g(1,u)Q\left(\frac{f(v)}{\sigma_w}\right)+g(0,u)Q\left(\frac{-f(v)}{\sigma_w}\right)\right)\Phi\left(\frac{v}{\sigma_v},\frac{u}{\sigma_u}\right)dudv\\
&\quad+\frac{\lambda}{\sigma_v}\int\Phi\left(\frac{v}{\sigma_v}\right) f(v)^2 dv\\\nonumber
&= \frac{-1}{\sigma_v}\int\left(v\int \frac{1}{\sigma_u} \Phi\left(\frac{v}{\sigma_v},\frac{u}{\sigma_u}\right)\left(\frac{r_1(u)}{r_2(u)}Q\left(\frac{f(v)}{\sigma_w}\right)+\frac{r_3(u)}{r_4(u)}Q\left(\frac{-f(v)}{\sigma_w}\right)\right) du\right.\\\label{Chptr4:Eqn_27}
&\left.\quad+\lambda \Phi\left(\frac{v}{\sigma_v}\right)f(v)^2\right)dv,
\end{align}
where
\begin{subequations}
\begin{align}
r_1(u)&\triangleq\int v\Phi\left(\frac{v}{\sigma_v},\frac{u}{\sigma_u}\right)Q\left(\frac{f(v)}{\sigma_w}\right)dv,\\
r_2(u)&\triangleq\int \Phi\left(\frac{v}{\sigma_v},\frac{u}{\sigma_u}\right)Q\left(\frac{f(v)}{\sigma_w}\right)dv,\\
r_3(u)&\triangleq\int v\Phi\left(\frac{v}{\sigma_v},\frac{u}{\sigma_u}\right)Q\left(\frac{-f(v)}{\sigma_w}\right)dv,\\
r_4(u)&\triangleq\int \Phi\left(\frac{v}{\sigma_v},\frac{u}{\sigma_u}\right)Q\left(\frac{-f(v)}{\sigma_w}\right)dv.
\end{align}
\end{subequations}
Note that (\ref{Chptr4:Eqn_27}) is in the form of (\ref{Chptr4:Eqn_18}) with $F(f,H(v),v)$ and $H(v)$ defined as
\begin{subequations}
\begin{align}
F(f,H(v),v)&=\frac{1}{\sigma_v} \left(-vH(v)+\lambda \Phi\left(\frac{v}{\sigma_v}\right)f(v)^2\right),\\
\text{and }H(v)&=\int G_0\left(f(v),r_1(u),\ldots,r_4(u),u,v\right) du,
\end{align}
\end{subequations}
where $G_0\left(f(v),r_1(u),\ldots,r_4(u),u,v\right),~G_i,i=1,...,4$ are given by
\begin{subequations}
\begin{align}
G_0\left(f(v),r_1(u),\ldots,r_4(u),u,v\right)&=\frac{1}{\sigma_u} \Phi\left(\frac{v}{\sigma_v},\frac{u}{\sigma_u}\right)\left(\frac{r_1(u)}{r_2(u)}Q\left(\frac{f(v)}{\sigma_w}\right)+\frac{r_3(u)}{r_4(u)}Q\left(\frac{-f(v)}{\sigma_w}\right)\right),\\
G_1&=v\Phi\left(\frac{v}{\sigma_v},\frac{u}{\sigma_u}\right)Q\left(\frac{f(v)}{\sigma_w}\right),\\
G_2&=\Phi\left(\frac{v}{\sigma_v},\frac{u}{\sigma_u}\right)Q\left(\frac{f(v)}{\sigma_w}\right),\\
G_3&=v\Phi\left(\frac{v}{\sigma_v},\frac{u}{\sigma_u}\right)Q\left(\frac{-f(v)}{\sigma_w}\right),\\
G_4&=\Phi\left(\frac{v}{\sigma_v},\frac{u}{\sigma_u}\right)Q\left(\frac{-f(v)}{\sigma_w}\right).
\end{align}
\end{subequations}
Now we can apply the necessary condition in (\ref{Chptr4:Eqn_19}). To this end, we compute
\begin{subequations}\label{Chptr4:Eqn_28}
\begin{align}
F^f(f(v),H(v),v)&=\frac{2\lambda}{\sigma_v} \Phi\left(\frac{v}{\sigma_v}\right)f(v),\\
F^H(f(v),H(v),v)&=\frac{-v}{\sigma_v}, \\
G_0^{f}(f(v),r_1(u),\ldots,r_4(u),u,v)&=\frac{e^{-\frac{f(v)^2}{2\sigma_w^2}}}{\sigma_w\sigma_u\sqrt{2\pi}} \Phi\left(\frac{v}{\sigma_v},\frac{u}{\sigma_u}\right)\left(\frac{r_3(u)}{r_4(u)}-\frac{r_1(u)}{r_2(u)}\right),\\
G_0^{r_1}(f(v),r_1(u),\ldots,r_4(u),u,v)&=\frac{1}{\sigma_ur_2(u)} \Phi\left(\frac{v}{\sigma_v},\frac{u}{\sigma_u}\right)Q\left(\frac{f(v)}{\sigma_w}\right),\\
G_0^{r_2}(f(v),r_1(u),\ldots,r_4(u),u,v)&=\frac{-r_1(u)}{\sigma_ur_2(u)^2} \Phi\left(\frac{v}{\sigma_v},\frac{u}{\sigma_u}\right)Q\left(\frac{f(v)}{\sigma_w}\right),\\
G_0^{r_3}(f(v),r_1(u),\ldots,r_4(u),u,v)&=\frac{1}{\sigma_ur_4(u)} \Phi\left(\frac{v}{\sigma_v},\frac{u}{\sigma_u}\right)Q\left(\frac{-f(v)}{\sigma_w}\right),\\
G_0^{r_4}(f(v),r_1(u),\ldots,r_4(u),u,v)&=\frac{-r_3(u)}{\sigma_ur_4(u)^2} \Phi\left(\frac{v}{\sigma_v},\frac{u}{\sigma_u}\right)Q\left(\frac{-f(v)}{\sigma_w}\right),\\
G_1^{f}(f(v),v,u)&= v\Phi\left(\frac{v}{\sigma_v},\frac{u}{\sigma_u}\right)\frac{- e^{-\frac{f(v)^2}{2\sigma_w^2}}}{\sqrt{2\pi}\sigma_w},\\
G_2^{f}(f(v),v,u)&= \Phi\left(\frac{v}{\sigma_v},\frac{u}{\sigma_u}\right)\frac{- e^{-\frac{f(v)^2}{2\sigma_w^2}}}{\sqrt{2\pi}\sigma_w},\\
G_3^{f}(f(v),v,u)&= v\Phi\left(\frac{v}{\sigma_v},\frac{u}{\sigma_u}\right)\frac{ e^{-\frac{f(v)^2}{2\sigma_w^2}}}{\sqrt{2\pi}\sigma_w},\\
G_4^{f}(f(v),v,u)&= \Phi\left(\frac{v}{\sigma_v},\frac{u}{\sigma_u}\right)\frac{ e^{-\frac{f(v)^2}{2\sigma_w^2}}}{\sqrt{2\pi}\sigma_w}.
\end{align}
\end{subequations}
Substituting (\ref{Chptr4:Eqn_28}) in (\ref{Chptr4:Eqn_19}), the necessary condition in (\ref{Chptr4:Eqn_18}) is obtained as
\begin{align}\nonumber
\nabla L&=\frac{2\lambda}{\sigma_v} \Phi\left(\frac{v}{\sigma_v}\right)f(v)-\frac{ve^{-\frac{f(v)^2}{2\sigma_w^2}}}{\sigma_v\sigma_w\sigma_u\sqrt{2\pi}}\int \Phi\left(\frac{v}{\sigma_v},\frac{u}{\sigma_u}\right)\left(\frac{r_3(u)}{r_4(u)}-\frac{r_1(u)}{r_2(u)}\right)du\\\nonumber
&\quad-\int\int\frac{t}{\sigma_v}\left(v\Phi\left(\frac{v}{\sigma_v},\frac{u}{\sigma_u}\right)\frac{- e^{-\frac{f(v)^2}{2\sigma_w^2}}}{\sqrt{2\pi}\sigma_w}\cdot \frac{1}{\sigma_ur_2(u)} \Phi\left(\frac{t}{\sigma_v},\frac{u}{\sigma_u}\right)Q\left(\frac{f(t)}{\sigma_w}\right)\right.\\\nonumber
&\quad+\Phi\left(\frac{v}{\sigma_v},\frac{u}{\sigma_u}\right)\frac{- e^{-\frac{f(v)^2}{2\sigma_w^2}}}{\sqrt{2\pi}\sigma_w}\cdot
\frac{-r_1(u)}{\sigma_ur_2(u)^2} \Phi\left(\frac{t}{\sigma_v},\frac{u}{\sigma_u}\right)Q\left(\frac{f(t)}{\sigma_w}\right)\\\nonumber
&\quad +v\Phi\left(\frac{v}{\sigma_v},\frac{u}{\sigma_u}\right)\frac{ e^{-\frac{f(v)^2}{2\sigma_w^2}}}{\sqrt{2\pi}\sigma_w}\cdot \frac{1}{\sigma_ur_4} \Phi\left(\frac{t}{\sigma_v},\frac{u}{\sigma_u}\right)Q\left(\frac{-f(t)}{\sigma_w}\right)\\\label{Chptr4:Eqn_29}
&\quad\left.+\Phi\left(\frac{v}{\sigma_v},\frac{u}{\sigma_u}\right)\frac{ e^{-\frac{f(v)^2}{2\sigma_w^2}}}{\sqrt{2\pi}\sigma_w}\cdot \frac{-r_3(u)}{\sigma_ur_4(u)^2} \Phi\left(\frac{t}{\sigma_v},\frac{u}{\sigma_u}\right)Q\left(\frac{-f(t)}{\sigma_w}\right)\right)dtdu=0.
\end{align}
Rewriting (\ref{Chptr4:Eqn_29}), we have
\begin{align}\nonumber
&2\sqrt{2\pi}\sigma_w\sigma_u\lambda\Phi\left(\frac{v}{\sigma_v}\right)f(v)e^{\frac{f(v)^2}{2\sigma_w^2}}=v\int \Phi\left(\frac{v}{\sigma_v},\frac{u}{\sigma_u}\right)\left(\frac{r_3(u)}{r_4(u)}-\frac{r_1(u)}{r_2(u)}\right)du\\\nonumber
&\quad- v\int\int t\Phi\left(\frac{v}{\sigma_v},\frac{u}{\sigma_u}\right)\cdot \frac{1}{r_2(u)} \Phi\left(\frac{t}{\sigma_v},\frac{u}{\sigma_u}\right)Q\left(\frac{f(t)}{\sigma_w}\right)dtdu\\\nonumber
&\quad+\int\int t\Phi\left(\frac{v}{\sigma_v},\frac{u}{\sigma_u}\right)\cdot
\frac{r_1(u)}{r_2(u)^2} \Phi\left(\frac{t}{\sigma_v},\frac{u}{\sigma_u}\right)Q\left(\frac{f(t)}{\sigma_w}\right)dtdu\\\nonumber
&\quad +v\int\int t\Phi\left(\frac{v}{\sigma_v},\frac{u}{\sigma_u}\right)\cdot \frac{1}{r_4(u)} \Phi\left(\frac{t}{\sigma_v},\frac{u}{\sigma_u}\right)Q\left(\frac{-f(t)}{\sigma_w}\right)dtdu\\
&\quad-\int\int t\Phi\left(\frac{v}{\sigma_v},\frac{u}{\sigma_u}\right)\cdot \frac{r_3(u)}{r_4(u)^2} \Phi\left(\frac{t}{\sigma_v},\frac{u}{\sigma_u}\right)Q\left(\frac{-f(t)}{\sigma_w}\right)dtdu\\\nonumber
&=v\int \Phi\left(\frac{v}{\sigma_v},\frac{u}{\sigma_u}\right)\left(\frac{r_3(u)}{r_4(u)}-\frac{r_1(u)}{r_2(u)}\right)du- v\int \Phi\left(\frac{v}{\sigma_v},\frac{u}{\sigma_u}\right)\cdot \frac{r_1(u)}{r_2(u)}du\\
&\quad+\int \Phi\left(\frac{v}{\sigma_v},\frac{u}{\sigma_u}\right)\cdot
\frac{r_1(u)^2}{r_2(u)^2}du+v\int \Phi\left(\frac{v}{\sigma_v},\frac{u}{\sigma_u}\right)\cdot \frac{r_3(u)}{r_4(u)}du-\int \Phi\left(\frac{v}{\sigma_v},\frac{u}{\sigma_u}\right)\cdot \frac{r_3(u)^2}{r_4(u)^2}du.
\end{align}
Finally, by some elementary manipulations the result in (\ref{Chptr4:Eqn_5}) is obtained.

\qed

\section{Proof of Proposition \ref{Chptr4:Prop_2} }\label{Chptr4:Appendix_4}
Assume that a decoder function $g(Y,U)$ is given. By expanding the Lagrangian function $L(f,g,\lambda)$ for the DOP we have
\begin{align}
L(f,g,\lambda)&=\epsilon(D)+\lambda \mathbb{E}[f(V)^2]\\\label{Chptr4:Eqn_30}
&=\frac{1}{\sigma_u}\int \epsilon (D|U=u)\Phi\left(\frac{u}{\sigma_u}\right)du+\frac{\lambda}{\sigma_v}\int\Phi\left(\frac{v}{\sigma_v}\right)f(v)^2dv.
\end{align}
Expanding $\epsilon (D|U=u)=\textrm{Pr}\left((V-\hat{V})^2\geq D|U=u\right)$ we have
\begin{subequations}
\begin{align}\nonumber
\epsilon(D|U=u)&=\textrm{Pr}(V\in I_0(u)\setminus I_1(u), Y=1|U=u)\\\nonumber
&\quad+\textrm{Pr}(V\in I_1(u)\setminus I_0(u), Y=0|U=u)\\\nonumber
&\quad+\textrm{Pr}(V\in (I_0(u)\cup I_1(u))^{C}, |\hat{V}-V|^2\geq D|U=u)\\
&\quad+\textrm{Pr}(V\in (I_0(u)\cap I_1(u)), |\hat{V}-V|^2\geq D|U=u)\\\nonumber
&=\frac{1}{\sigma_v}\!\!\!\!\!\!\int\limits_{v\in I_0(u)\setminus I_1(u)}\!\!\!\!\!\!\Phi\left(\frac{v}{\sigma_v}\Big|\frac{u}{\sigma_u}\right)Q\left(\frac{f(v)}{\sigma_w}\right)dv\\\nonumber
&\quad+\frac{1}{\sigma_v}\!\!\!\!\!\!\int\limits_{v\in I_1(u)\setminus I_0(u)}\!\!\!\!\!\!\Phi\left(\frac{v}{\sigma_v}\Big|\frac{u}{\sigma_u}\right)Q\left(\frac{-f(v)}{\sigma_w}\right) dv\\\label{Chptr4:Eqn_31}
&\quad+\frac{1}{\sigma_v}\!\!\!\!\!\!\int\limits_{v\in (I_0(u)\cup I_1(u))^C}\!\!\!\!\!\!\Phi\left(\frac{v}{\sigma_v}\Big|\frac{u}{\sigma_u}\right)dv,
\end{align}
\end{subequations}
where we used the fact that no outage occurs when $V\in I_0(U)\cap I_1(U)$. Substituting (\ref{Chptr4:Eqn_31}) in (\ref{Chptr4:Eqn_30}), we can write the Lagrangian $L(f,g,\lambda)$ as
\begin{align}
L(f,g,\lambda)&=\frac{1}{\sigma_v\sigma_u}\int \Phi\left(\frac{u}{\sigma_u}\right)\int \Phi\left(\frac{v}{\sigma_v}\Big|\frac{u}{\sigma_u}\right)G\left(u,v,f(v)\right)dvdu+\frac{\lambda}{\sigma_v}\int \Phi\left(\frac{v}{\sigma_v}\right)\lambda f^2(v)dv\\\label{Chptr4:Eqn_32}
&=\frac{1}{\sigma_v}\int \Phi\left(\frac{v}{\sigma_v}\right)\left[\int \frac{1}{\sigma_u}\Phi\left(\frac{u}{\sigma_u}\Big|\frac{v}{\sigma_v}\right)G\left(u,v,f(v)\right)du+\lambda f^2(v)\right]dv,
\end{align}
with $G\left(u,v,f(v)\right)$ defined as
\begin{align}
G\left(u,v,f(v)\right)&\triangleq\left\{\begin{array}{ll}
Q\left(\frac{f(v)}{\sigma_w}\right)& v\in (I_0(u)\setminus I_1(u)), \\
Q\left(\frac{-f(v)}{\sigma_w}\right)& v\in (I_1(u)\setminus I_0(u)),\\
1& v\in (I_0(u)\cup I_1(u))^{C},\\
0& v\in (I_0(u)\cap I_1(u)).
\end{array}
\right.
\end{align}
Note that (\ref{Chptr4:Eqn_32}) is in the form of (\ref{Chptr4:Eqn_18}) with $F(f,H(v),v)$ and $H(v)$ given by
\begin{subequations}
\begin{align}
F(f,H(v),v)&=\frac{1}{\sigma_v}\Phi\left(\frac{v}{\sigma_v}\right)\cdot\left(H(v)+\lambda f^2(v)\right),\\
H(v)&=\int G_0\left(f(v),u,v\right)du,
\end{align}
\end{subequations}
respectively, where $G_0\left(f(v),u,v\right)$ is given by
\begin{align}
G_0\left(f(v),u,v\right)&=\frac{1}{\sigma_u}\Phi\left(\frac{u}{\sigma_u}\Big|\frac{v}{\sigma_v}\right)G\left(u,v,f(v)\right).
\end{align}
Applying the necessary condition in (\ref{Chptr4:Eqn_19}) for the optimal solution, for different terms in (\ref{Chptr4:Eqn_19}) we have
\begin{subequations}
\begin{align}
F^f(f(v),H(v),v)&=\frac{2\lambda}{\sigma_v} \Phi\left(\frac{v}{\sigma_v}\right)f(v),\\
F^H(f(v),H(v),v)&=\frac{1}{\sigma_v}\Phi\left(\frac{v}{\sigma_v}\right), \\
G_0^{f}(f(v),u,v)&=\frac{1}{\sigma_u}\Phi\left(\frac{u}{\sigma_u}\Big|\frac{v}{\sigma_v}\right)G^f\left(u,v,f(v)\right),
\end{align}
\end{subequations}
where $G^{f}\left(u,v,f(v)\right)$ is obtained as
\begin{align}\label{Chptr4:Eqn_33}
G^f\left(u,v,f(v)\right)&=\left\{\begin{array}{ll}
\frac{-1}{\sqrt{2\pi}}e^{-\frac{f(v)^2}{2\sigma_w^2}}& v\in (I_0(u)\setminus I_1(u)) \\
\frac{1}{\sqrt{2\pi}}e^{-\frac{f(v)^2}{2\sigma_w^2}}& v\in (I_1(u)\setminus I_0(u))\\
0& v\in (I_0(u)\cap I_1(u))~\text{or}~v\in (I_0(u)\cup I_1(u))^{C}
\end{array}.
\right.
\end{align}
Therefore, (\ref{Chptr4:Eqn_19}) can be written as
\begin{align}\label{Chptr4:Eqn_34}
\nabla L=\frac{1}{\sigma_v} \Phi\left(\frac{v}{\sigma_v}\right)\left(2\lambda f(v)+\frac{1}{\sigma_u}\int \Phi\left(\frac{u}{\sigma_u}\Big|\frac{v}{\sigma_v}\right)G^f\left(u,v,f(v)\right)du\right)=0.
\end{align}

Note that the integration in (\ref{Chptr4:Eqn_34}) is over the side information $u$. In the following, we aim at identifying the boundaries of the intervals of $u$, such that, for a given source output $v$ we have $G^f\left(u,v,f(v)\right)\neq0$. To do so, we characterize the intervals as
\begin{align}\nonumber
I_0(u)&\setminus I_1(u)=\left(b_{0l}(u),b_{0r}(u)\right),\\
I_1(u)&\setminus I_0(u)=\left(b_{1l}(u),b_{1r}(u)\right).
\end{align}
Note that if $v\in I_0(u)\cap I_1(u)$ and $v\in (I_0(u)\cup I_1(u))^{C}$, we have $G^f\left(u,v,f(v)\right)=0$. For a given side information realization $u$, $g(0,u)$ and $g(1,u)$ are two points. Hence, depending on the condition that $g(0,u)$ is equal to, less than, or greater than $g(1,u)$, we have different situations for $I_0(u)$ and $I_1(u)$ in (\ref{Chptr4:Eqn_33}).

\textit{Case 1) $g(0,u)=g(1,u)$}: In this case the two intervals $I_0(u)$ and $I_1(u)$ overlap completely, and therefore, $I_0(u)\setminus I_1(u)$ and $I_1(u)\setminus I_0(u)$ are both empty sets.

\textit{Case 2) $g(0,u)>g(1,u)$}: In this case $b_{0l}(u)$, $b_{0r}(u)$, $b_{1r}(u)$ and $b_{1l}(u)$ are obtained as
\begin{align}\nonumber
b_{0r}(u)&=g(0,u)+\sqrt{D},\\\nonumber
b_{0l}(u)&=\max\left\{g(1,u)+\sqrt{D},g(0,u)-\sqrt{D}\right\},\\\nonumber
b_{1r}(u)&=\min\left\{g(1,u)+\sqrt{D},g(0,u)-\sqrt{D}\right\},\\
b_{1l}(u)&=g(1,u)-\sqrt{D}.
\end{align}

\textit{Case 3) $g(0,u)<g(1,u)$}: In this case $b_{0l}(u)$, $b_{0r}(u)$, $b_{1r}(u)$ and $b_{1l}(u)$ are obtained as
\begin{align}\nonumber
b_{0r}(u)&=\min\left\{g(0,u)+\sqrt{D},g(1,u)-\sqrt{D}\right\},\\\nonumber
b_{0l}(u)&=g(0,u)-\sqrt{D},\\\nonumber
b_{1r}(u)&=g(1,u)+\sqrt{D},\\
b_{1l}(u)&=\max\left\{g(1,u)-\sqrt{D},g(0,u)+\sqrt{D}\right\}.
\end{align}
It can be easily verified that for a given source output $v$, the side information range corresponding to $G^f\left(u,v,f(v)\right)\neq0$ can be obtained as $S_{0\setminus1}(v)\cup S_{1\setminus0}(v)$, where $S_{0\setminus1}(v)$ and $S_{1\setminus0}(v)$ are defined as
\begin{align}\nonumber
S_{0\setminus1}(v)&\triangleq\{u:~b_{0r}(u)\geq v\geq b_{0l}(u)\},\\
S_{1\setminus0}(v)&\triangleq\{u:~b_{1r}(u)\geq v\geq b_{1l}(u)\}.
\end{align}
Finally, we can simplify (\ref{Chptr4:Eqn_34}) as
\begin{align}\nonumber
\nabla L &=\frac{1}{\sigma_v}\Phi\left(\frac{v}{\sigma_v}\right)\left[\int\limits_{u\in S_{0\setminus1}(v)}\frac{-1}{\sqrt{2\pi}\sigma_u}e^{-\frac{f(v)^2}{2\sigma_w^2}} \Phi\left(\frac{u}{\sigma_u}\Big|\frac{v}{\sigma_v}\right)du\right.\\
&\left.\quad\quad\quad\quad\quad+\int\limits_{u\in S_{1\setminus0}(v)}\frac{1}{\sqrt{2\pi}\sigma_u}e^{-\frac{f(v)^2}{2\sigma_w^2}} \Phi\left(\frac{u}{\sigma_u}\Big|\frac{v}{\sigma_v}\right)du+2\lambda f(v)\right]\\\nonumber
&=\frac{1}{\sigma_v}\Phi\left(\frac{v}{\sigma_v}\right)\left[\frac{-1}{\sqrt{2\pi}\sigma_u}e^{-\frac{f(v)^2}{2\sigma_w^2}} \int\limits_{u\in S_{0\setminus1}(v)}\Phi\left(\frac{u}{\sigma_u}\Big|\frac{v}{\sigma_v}\right)du\right.\\\label{Chptr4:Eqn_35}
&\left.\quad\quad\quad\quad\quad+\frac{1}{\sqrt{2\pi}\sigma_u}e^{-\frac{f(v)^2}{2\sigma_w^2}} \int\limits_{u\in S_{1\setminus0}(v)} \Phi\left(\frac{u}{\sigma_u}\Big|\frac{v}{\sigma_v}\right)du+2\lambda f(v)\right].
\end{align}
Imposing (\ref{Chptr4:Eqn_35}) to be zero we have
\begin{align}
f(v)&=\frac{e^{-\frac{f(v)^2}{2\sigma_w^2}}}{2\lambda\sqrt{2\pi}}\left(\textrm{Pr}\left(U\in S_{0\setminus1}(v)\right)-\textrm{Pr}\left(U\in S_{1\setminus0}(v)\right)\right).
\end{align}
\qed

\section{Proof of Proposition \ref{Chptr4:Prop_3} }\label{Chptr4:Appendix_5}

The optimal decoder functions, i.e., $g(0,u)$ and $g(1,u)$ can be obtained as
\begin{align}
g(0,u)& =\stackunder[5pt]{arg min}{$\hat{v}$}~~\textrm{Pr}\left(|V-\hat{v}|^2\geq D|U=u, Y=0\right)\\
&=\stackunder[5pt]{arg max}{$\hat{v}$}~~\textrm{Pr}\left(|V-\hat{v}|^2< D|U=u, Y=0\right)\\
&=\stackunder[5pt]{arg max}{$\hat{v}$}~~\frac{1}{\sigma_v\sigma_u}\int\limits_{\hat{v}-\sqrt{D}}^{\hat{v}+\sqrt{D}}p_{V|U,Y}\left(t|u,Y=0\right)dt\\\label{Chptr4:Eqn_37}
&=\stackunder[5pt]{arg max}{$\hat{v}$}~~\int\limits_{\hat{v}-\sqrt{D}}^{\hat{v}+\sqrt{D}}\Phi\left(\frac{t}{\sigma_v}\Big|\frac{u}{\sigma_u}\right)Q\left(\frac{-f(t)}{\sigma_w}\right)dt.
\end{align}
We note that, since the mapping $f(v)$ is given, it could be possible that for some encoder mapping $f$ and side information realization $u$, more than one output is obtained in (\ref{Chptr4:Eqn_37}). From the DOP point of view, there is no difference in choosing either of these points. Therefore, we have
\begin{align}
g^{*}(0,u)&\in~~\stackunder[5pt]{arg max}{$\hat{v}$}~~\int\limits_{\hat{v}-\sqrt{D}}^{\hat{v}+\sqrt{D}}\Phi\left(\frac{t}{\sigma_v},\frac{u}{\sigma_u}\right)Q\left(\frac{-f(t)}{\sigma_w}\right)dt,
\end{align}
and similarly for $g(1,u)$, we have
\begin{align}
g^{*}(1,u)&\in~~\stackunder[5pt]{arg max}{$\hat{v}$}~~\int\limits_{\hat{v}-\sqrt{D}}^{\hat{v}+\sqrt{D}}\Phi\left(\frac{t}{\sigma_v},\frac{u}{\sigma_u}\right)Q\left(\frac{f(t)}{\sigma_w}\right)dt.
\end{align}

\qed

\section{Proof of (\ref{Chptr4:Eqn_41}) in Remark \ref{Chptr4:Remark_4} }\label{Chptr4:Appendix_6}
In the low SNR regime (large values of $\sigma_w^2$), it can be verified from (\ref{Chptr4:Eqn_12}) that the encoder mapping tends to an all-zero function. Hence, the DOP can be evaluated as
\begin{subequations}
\begin{align}
  \epsilon(D) &= 1-\textrm{Pr}(|V-\hat{V}|^2< D) \\
  &=1-\frac{1}{\sigma_u}\int \textrm{Pr}(|V-\hat{V}|^2< D|U=u)\Phi\left(\frac{u}{\sigma_u}\right)du\\
  &=1-\frac{1}{\sigma_u}\int\Phi\left(\frac{u}{\sigma_u}\right) \textrm{Pr}\left(\big|V-\frac{r\sigma_v}{\sigma_u}u\big|^2< D|U=u\right)du\\
  &=1-\frac{1}{\sigma_u\sigma_v}\int\Phi\left(\frac{u}{\sigma_u}\right) \int\limits_{\frac{r\sigma_v}{\sigma_u}u-\sqrt{D}}^{\frac{r\sigma_v}{\sigma_u}u+\sqrt{D}}\Phi\left(\frac{v}{\sigma_v}\Big|\frac{u}{\sigma_u}\right)du\\
  &=1-\frac{1}{\sigma_u}\int\Phi\left(\frac{u}{\sigma_u}\right) \left(Q\left(\frac{-\sqrt{D}}{\sigma_v\sqrt{1-r^2}}\right)-Q\left(\frac{\sqrt{D}}{\sigma_v\sqrt{1-r^2}}\right)\right)du\\
  &=2Q\left(\frac{\sqrt{D}}{\sigma_v\sqrt{1-r^2}}\right).
\end{align}
\end{subequations}

\qed

\section{Proof of Proposition \ref{Chptr4:Prop_4} }\label{Chptr4:Appendix_7}
We first define as $f(\cdot)$ the encoder mapping applied to the error in (\ref{Chptr4:Eqn_8}). Assuming that the side information is available at both the encoder and the decoder, the optimal decoder can be obtained as
\begin{subequations}
\begin{align}
g(y,u)&=\arg \min_{\hat{v}}~~ \textrm{Pr}\left(|V-\hat{v}|^2\geq D|Y=y,U=u\right)+\lambda \mathbb{E}[\tilde{f}(T)^2]\\
&=\arg \min_{\hat{v}}~~\int\limits_{\hat{v}-\sqrt{D}}^{\hat{v}+\sqrt{D}}\Phi\left(\frac{v}{\sigma_v},\frac{u}{\sigma_u}\right)Q\left(\frac{(-1)^{y+1}\tilde{f}\left(v-\frac{r\sigma_v u}{\sigma_u}\right)}{\sigma_w}\right)dv-\lambda \mathbb{E}[\tilde{f}(T)^2]\\
&=\arg \min_{\hat{v}}~~\int\limits_{\hat{v}-\sqrt{D}}^{\hat{v}+\sqrt{D}}\Phi\left(\frac{v}{\sigma_v}\Big|\frac{u}{\sigma_u}\right)Q\left(\frac{(-1)^{y+1}\tilde{f}\left(v-\frac{r\sigma_v u}{\sigma_u}\right)}{\sigma_w}\right)dv-\lambda \mathbb{E}[\tilde{f}(T)^2]\\
&=\arg \min_{\hat{v}}~~\int\limits_{\hat{v}-\sqrt{D}}^{\hat{v}+\sqrt{D}}e^{-\frac{\left(v-\frac{r\sigma_v u}{\sigma_u}\right)^2}{2\sigma_v^2(1-r^2)}}Q\left(\frac{(-1)^{y+1}\tilde{f}\left(v-\frac{r\sigma_v u}{\sigma_u}\right)}{\sigma_w}\right)dv-\lambda \mathbb{E}[\tilde{f}(T)^2]\\\label{Chptr4:Eqn_46}
&=\arg \min_{\hat{v}}~~\int\limits_{\hat{v}-\sqrt{D}-\frac{r\sigma_v u}{\sigma_u}}^{\hat{v}+\sqrt{D}-\frac{r\sigma_v u}{\sigma_u}}e^{-\frac{t^2}{2\sigma_v^2(1-r^2)}}Q\left(\frac{(-1)^{y+1}\tilde{f}(t)}{\sigma_w}\right)dt-\lambda \mathbb{E}[\tilde{f}(T)^2]\\\label{Chptr4:Eqn_47}
&=\frac{r\sigma_v u}{\sigma_u}+\stackunder[5pt]{arg min}{$\hat{t}$}~~\int\limits_{\hat{t}-\sqrt{D}}^{\hat{t}+\sqrt{D}}e^{-\frac{t^2}{2\sigma_v^2(1-r^2)}}Q\left(\frac{(-1)^{y+1}\tilde{f}(t)}{\sigma_w}\right)dt-\lambda \mathbb{E}[\tilde{f}(T)^2]\\\label{Chptr4:Eqn_48}
&=\frac{r\sigma_v u}{\sigma_u}+\hat{t}_Y,
\end{align}
\end{subequations}
where in (\ref{Chptr4:Eqn_46}) we used the transformation $v-\frac{r \sigma_v u}{\sigma_u}=t$; in (\ref{Chptr4:Eqn_47}) we replaced $\hat{v}-r\sigma_v u/\sigma_u$ with $\hat{t}$ by adding $r\sigma_v u/\sigma_u$ to the resultant argument. Finally, the second term in (\ref{Chptr4:Eqn_48}) represent the optimal decoder when there is no side information as derived in \cite[Proposition IV.2]{arive_varsteh_rassouli_osvaldo_gunduz}.

The DOP in (\ref{Chptr4:Eqn_45}) can be evaluated as
\begin{subequations}
\begin{align}
\epsilon(D)&=\frac{1}{\sigma_v}\int \Phi\left(\frac{u}{\sigma_u}\right)\textrm{Pr}\left(|V-\hat{V}|^2\geq D|U=u\right)du\\
&=\frac{1}{\sigma_v}\int \Phi\left(\frac{u}{\sigma_u}\right)\textrm{Pr}\left(\Big|V-\frac{r\sigma_v u}{\sigma_u}-\hat{t}_Y\Big|^2\geq D\Big|U=u\right)du\\
&=\frac{1}{\sigma_v}\int \Phi\left(\frac{u}{\sigma_u}\right)\Big[\textrm{Pr}\left(V\in\left(I_0(u)\cup I_1(u)\right)^C\right)+\\
&\quad\quad\textrm{Pr}\left(V\in I_0(u)\setminus I_1(u), \hat{t}_Y=t_1\right)+\textrm{Pr}\left(V\in I_1(u)\setminus I_0(u), \hat{t}_Y=t_0\right)\Big]du\\
&=\frac{1}{\sigma_u\sigma_v}\int \Phi\left(\frac{u}{\sigma_u}\right)\left[\int\limits_{(I_0(u)\cup I_1(u))^C}\Phi\left(\frac{v}{\sigma_v}\Big|\frac{u}{\sigma_u}\right)dv+\right.\\
&\quad\quad\left.Q\left(\frac{t}{\sigma_w}\right)\left(\int\limits_{I_1(u)\setminus I_0(u)}\Phi\left(\frac{v}{\sigma_v}\Big|\frac{u}{\sigma_u}\right)dv+\int\limits_{I_0(u)\setminus I_1(u)}\Phi\left(\frac{v}{\sigma_v}\Big|\frac{u}{\sigma_u}\right)dv\right)\right],\\
\end{align}
\end{subequations}
where we have defined
\begin{align}
I_y(u)\triangleq\left\{v:\left(v-\frac{r\sigma_v u}{\sigma_u}-\hat{t}_y\right)^2\leq D\right\},~y=0,1.
\end{align}

\qed

\bibliographystyle{ieeetran}
\bibliography{ref}

\end{document}